\documentclass[12pt]{article}
\pdfoutput=1

\usepackage{draft} 
\usepackage{hyperref}
\usepackage{graphicx,color}

\begin{document}

\unitlength = .8mm

\begin{titlepage}
\rightline{MIT-CTP-4639}

\begin{center}

\hfill \\
\hfill \\
\vskip 1cm


\title{Interpolating the Coulomb Phase of Little String Theory}

\author{Ying-Hsuan Lin$^\spadesuit$, Shu-Heng Shao$^\spadesuit$, Yifan Wang$^\diamondsuit$, Xi Yin$^\spadesuit$}

\address{$^\spadesuit$Jefferson Physical Laboratory, Harvard University, \\
Cambridge, MA 02138 USA
\\
$^\diamondsuit$Center for Theoretical Physics, Massachusetts Institute of Technology, \\
Cambridge, MA 02139 USA}

\email{yhlin@physics.harvard.edu, shshao@physics.harvard.edu, \\ yifanw@mit.edu,
xiyin@fas.harvard.edu}

\end{center}

\abstract{ We study up to 8-derivative terms in the Coulomb branch effective action of $(1,1)$ little string theory, by collecting results of 4-gluon scattering amplitudes from both perturbative 6D super-Yang-Mills theory up to 4-loop order, and tree-level double scaled little string theory (DSLST). 
In previous work we have matched the 6-derivative term from the 6D gauge theory to DSLST, indicating that this term is protected on the entire Coulomb branch. The 8-derivative term, on the other hand, is unprotected.
 In this paper we compute the 8-derivative term by interpolating from the two limits, near the origin and near the infinity on the Coulomb branch, numerically from $SU(k)$ SYM and DSLST respectively, for $k=2,3,4,5$. We discuss the implication of this result on the UV completion of 6D SYM as well as the strong coupling completion of DSLST.
We also comment on analogous interpolating functions in the Coulomb phase of circle-compactified $(2,0)$ little string theory.
}

\vfill

\end{titlepage}

\eject

\tableofcontents

\section{Introduction}

The $(1,1)$ $A_{k-1}$ little string theory (LST) \cite{Berkooz:1997cq, Seiberg:1997zk, Aharony:1998ub,Giveon:1999zm,Aharony:1999ks,Kutasov:lecturenote} may be thought of as a UV completion of the 6-dimensional maximally supersymmetric $SU(k)$ Yang-Mills theory. The double scaled little string theory (DSLST) \cite{Giveon:1999px, Giveon:1999tq} is a particularly useful deformation of LST that admits a perturbative expansion, and describes the Coulomb phase of the 6D gauge theory far from the origin on the Coulomb branch. The perturbative description of the gauge theory, on the other hand, may be regarded as an expansion near the origin of the Coulomb branch, and describes the strong coupling limit of DSLST. The goal of this paper is to exploit this correspondence, by connecting the two limits of the Coulomb phase of $(1,1)$ LST.

We will inspect the derivative expansion of the Coulomb branch effective action, focusing on terms of the structure $f_n(r) D^{2n} F^4$, $n=0,1,2,$ etc. Here $r$ stands for the distance from the origin of the Coulomb branch, as measured by the scalar expectation values, and $F$ the field strength of the $U(1)^{k-1}$ vector multiplets in the Cartan of the $SU(k)$ gauge group. The most convenient way to organize the supersymmetric completion of these higher derivative terms in the effective action is through the massless superamplitudes they generate \cite{Elvang:2010xn}. For our purpose, it suffices to focus on the 4-point superamplitudes, which take the form\footnote{For comparison, the color-ordered tree-level superamplitude  is given by $\mathcal{A}^{tree} = -{i\over s_{12}s_{14}} \delta^8 (Q)$. However, note that when the external gluon states are restricted to the Cartan subalgebra, the tree amplitude vanishes identically.} $\delta^8(Q) F(s,t,u)$, where $Q$ is the total supermomentum and $s,t,u$ the Mandelstam variables \cite{Cheung:2009dc,Dennen:2009vk,Bern:2010qa}. $F(s,t,u)$ will depend on the color assignment of the Cartan gluons, and depend on $r$ through the $W$-boson masses.

The 4-point superamplitude can be computed in the large $r$ regime by the perturbative double scaled LST \cite{Aharony:2003vk,Chang:2014jta}. In previous work we have formulated the tree amplitude in the DSLST in terms of an explicit double integration over the cross ratio of four points on the Riemann sphere and over a continuous family of conformal blocks, which is then evaluated numerically. In this paper we will present some higher order terms in the $\A'$-expansion of the DSLST tree amplitude, giving the leading $1/r^2$ term of the $f_n(r) D^{2n} F^4$ coupling on the Coulomb branch, at large $r$.

In the small $r$ regime, on the other hand, we will perform a perturbative computation in 6D $SU(k)$ SYM. The 4-point amplitude is reduced to $\delta^8(Q)$ times a set of scalar box type integrals, which can be evaluated straightforwardly up to 3-loops. We will present some numerical results for $k=2,3,4,5$. Starting at 4-loop order, the 4-point amplitude of Cartan gluons suffers from logarithmic UV divergences. This divergence structure is a bit intricate, as the {\it non-abelian} 4-point amplitude already diverges at 3-loop and a 3-loop counter-term of the form $D^2 {\rm tr} F^4$ is needed \cite{Bern:2010tq,Bern:2012uf}. While this counter-term vanishes when restricted to the Cartan, it gives a nontrivial contribution to the 4-loop amplitude, which has been studied in \cite{Bern:2012uf}. In the end, after taking into account suitable 4-loop counter-terms, of the form $D^4 {\rm tr} F^4$ and $D^4{\rm tr}^2 F^4$, one obtains a 4-loop contribution to $f_2(r)$ that involves logarithmic dependence on $r$, of the form $(\ln r)^2$ and $\ln r$. While the {\it finite} shifts of the 3-loop and 4-loop counter-terms are not a priori determined in SYM perturbation theory (but should be ultimately fixed in the LST), the coefficients of the leading logarithms are unambiguously determined. The results of \cite{Bern:2012uf} on the 4-loop divergence of double trace terms then allows for determining certain leading log coefficients, which when combined with $1,2,3$-loop results produce the first few terms in the small $r$ expansion of $f_n(r)$.

The agreement of the $r^{-2} F^4$ term between a 1-loop computation of 6D SYM and low energy limit of DSLST found in \cite{Aharony:2003vk}, was expected as a consequence of the supersymmetry constraints on the $F^4$ coupling in the Coulomb branch effective action \cite{Seiberg:1997zk,Paban:1998ea,Sethi:1999qv}. The agreement of $r^{-2} D^2 F^4$ term between a 2-loop computation of 6D SYM, the next order $\A'$-expansion of the DSLST amplitude was found in our previous work \cite{Chang:2014jta}, numerically for $k=2,3,4,5$. One anticipates that this agreement should follow from supersymmetry constraints on $D^2 F^4$ coupling, namely the function $f_1(r)$ should be fixed to be the form $C_1/r^2$, and the coefficient $C_1$ can then be computed from either small $r$ (SYM) or large $r$ (DSLST). Indeed, the agreement we found in the $SU(3)$ case can be understood in terms of the (sixteen-supercharge) non-renormalization theorem of \cite{Sethi:1999qv}\footnote{For $SU(2)$ gauge theory, the $D^2 F^4$ term in the Lagrangian is proven to be two-loop exact by \cite{Paban:1998qy, Maxfield:2012aw}. But this is essentially equivalent to the statement that there is no nontrivial independent $D^2 F^4$ coupling in the Coulomb effective action of the $SU(2)$ theory, as the corresponding four-Cartan gluon superamplitude vanishes trivially.}. Although the result of \cite{Sethi:1999qv} is not directly applicable to $k>3$, we expect a similar non-renormalization theorem to hold for general $k$.


The focus of this paper is the $f_2(r) D^4 F^4$ term. This is the lowest order in the derivative expansion of the Coulomb branch effective action where we anticipate a nontrivial interpolating function $f_2(r)$ from small $r$ (SYM) to large $r$ (DSLST). Indeed, $f_2(r)$ receives all loop perturbative contributions. Collecting numerical results on both sides, we will be able to estimate the interpolating function on the entire Coulomb branch. We will find that, while the small and large $r$ limits are obviously different expansions, when naively extrapolated to the intermediate regime they are not far from one another. 

In the next section, we describe the general structure of the Coulomb branch effective action and its relation to superamplitudes. Then we will describe the perturbative computation of up to 8-derivative terms in the Coulomb branch effective action, from up to 4-loop results in the gauge theory. In section 4, we collect the results from DSLST tree amplitude, expanded to the appropriate orders in $\A'$. We then inspect numerically $f_2(r) D^4 F^4$ on the entire Coulomb branch, from small to large $r$. Implications of this result on the UV completion of perturbative 6D SYM, as well as the strong coupling completion of perturbative DSLST, will be discussed. 

Finally, in section 6, we will discuss the compactification of the $(2, 0)$ LST to five dimensions, and constrain the resulting 5D gauge theory by considerations of the effective action in the Coulomb phase of the compactified $(2,0)$ LST.

\section{The Coulomb Branch Effective Action}

The Coulomb branch moduli space of the $A_{k-1}$ LST is $(\mathbb{R}^4)^{k-1}/S_k$, parameterized by the value of $4(k-1)$ massless scalars in 6 dimensions \cite{Aharony:1998ub}. We denote these massless scalar by $\phi_i$, $i=1,2,3,4$, which take values in the $U(1)^{k-1}$ Cartan of the $SU(k)$ gauge group, in the 6D SYM description (which is a priori valid near the origin of the Coulomb branch). We will focus on a $\mathbb{Z}_k$-invariant 1-dimensional subspace of the Coulomb moduli space, corresponding to 
\ie\label{zphi}
& Z \equiv \phi_1 + i\phi_2 = r\, {\rm diag} ( 1,e^{2\pi i/k},\cdots, e^{2\pi i(k-1)/k} ),
\\
& \phi_3=\phi_4=0.
\fe
The large $r$ regime along this 1-dimensional subspace is then described by the perturbative double scaled little string theory \cite{Giveon:1999px,Giveon:1999tq}, with the worldsheet CFT given by
\ie
\mathbb{R}^{1,5}\times {(SL(2)_k/U(1)) \times (SU(2)_k/U(1)) \over \mathbb{Z}_k}.
\fe
The string coupling at the tip of the cigar (target space of $SL(2)/U(1)$ coset CFT) is identified with $1/r$.

The massless degrees of freedom in the Coulomb phase, consisting of $k-1$ Abelian vector multiplets of the 6D $(1,1)$ supersymmetry, are governed by a quantum effective action, that is the $U(1)^{k-1}$ supersymmetric gauge theory action together with an infinite series of higher derivative couplings. We will focus on couplings of the schematic form $f(\phi) D^{2n} F^4+\cdots$. Such higher derivative deformations of the Abelian $(1,1)$ gauge theory are constrained by supersymmetry, though the constraints become weaker with increasing number of derivatives. An illuminating way to organize the higher derivative couplings is through the corresponding supervertex, namely, a set of (super)amplitudes that obey supersymmetry Ward identities with no poles \cite{Elvang:2010xn}. If we fix the scalar vev (say of the form (\ref{zphi})), and consider terms of the form $D^{2n} F^4+\cdots$, then a supersymmetric completion of such a coupling corresponds to a 4-point supervertex of the form
\ie
\delta^8(Q) F(s,t,u),
\fe
where $Q$ is the total supermomentum, defined by \cite{Cheung:2009dc,Dennen:2009vk,Bern:2010qa}
\ie
& Q = \sum_{i=1}^4 {\bf q}_i,~~~~ {\bf q_i} = (q_i^A, \widetilde q_{iB}),
\\
& q_i^A = \lambda_i^{Aa} \eta_{ia}, ~~~\widetilde q_{iB} = \widetilde\lambda_{iB\dot b} \widetilde\eta^{\dot b}_i.
\fe
Here $i$ labels the external lines of the amplitude, $A,B=1,\cdots,4$ are $SO(1,5)$ Lorentz spinor indices, $a$ and $\dot b$ on the other hand are $SU(2)\times SU(2)$ little group indices. $\lambda_i^{Aa}$ and $\widetilde\lambda_{iB\dot b}$ are 6 dimensional spinor helicity variables, with the null momentum of the $i$-th particle related by $p_i^{AB} = \lambda_i^{Aa}\lambda_i^{Bb}\epsilon_{ab}$, $p_{iAB} = \widetilde\lambda_{iA\dot a} \widetilde\lambda_{iB\dot b}\epsilon^{\dot a\dot b} = {1\over 2}\epsilon_{ABCD}p_i^{CD}$. $\eta_{ia}$ and $\widetilde \eta^\db_i$ are a set of 4 Grassmannian variables that generate the $2^4=16$ states in the supermultiplet of the $i$-th particle.

Corresponding to $D^{2n}F^4$ coupling, $F(s,t,u)$ would be a function of Mandelstam variables $s,t,u$ of total degree $n$. For instance, if we fix the color structure (choice of Cartan generators), there is a unique supersymmetric completion of the $F^4$ term, corresponding to the constant term in $F(s,t,u)$. In the $SU(2)$ gauge theory, the massless fields on the Coulomb branch are in a single $U(1)$ gauge multiplet, and thus $F(s,t,u)$ must be symmetric in $s,t,u$. From this we immediately learn that there is no independent $D^2 F^4$ vertex, since $s+t+u=0$. This result is also an immediate consequence of the non-renormalization theorem of Paban, Sethi, and Stern \cite{Paban:1998qy} which is later extended to the $SU(3)$ case by \cite{Sethi:1999qv}. In the more general $SU(k)$ theory with $k>3$, to the best of our knowledge, there isn't a non-renormalization theorem that determines the $D^2F^4$ completely in terms of the $F^4$ coupling on the Coulomb branch. In fact, since different Cartan generators can be assigned to the 4 external lines of the superamplitude, one can construct nontrivial superamplitudes with $F(s,t,u)$ a linear function of $s,t,u$. These are the terms computed in \cite{Chang:2014jta}, from both the SYM at 2-loop and from DSLST. It is likely that by consideration of higher point superamplitudes, and consistency with unitarity, one can derive the supersymmetry constraint on the $r$-dependence of the $f_1(r) D^2 F^4$ coupling as in the work of Sethi, but we not will pursue this topic in the current paper.

The consideration of superamplitudes allows for an easy classification of $D^{2n}F^4$ couplings for all $n$. In below we will mostly think in terms of the superamplitudes rather than the terms in the effective Lagrangian. Now to be precise we will introduce a color label $a_i\in\mathbb{Z}_k$ for each external line, corresponding to a Cartan gluon in the $U(1)^{k-1}$ that transforms under the $\mathbb{Z}_k$ cyclic permutation of $k$ NS5-branes by the phase $e^{2\pi i a_i/k}$. The 4-point superamplitude is of course subject to the constraint $\sum_{i=1}^4 a_i=0$ (mod $k$), and takes the form
\ie
\delta^8(Q) F_{a_1a_2a_3a_4}(s,t,u; r),
\fe
where our convention, $s=s_{12}=-(p_1+p_2)^2$, $t=s_{14}$, $u=s_{13}=-s-t$. We also have the following identification between the 6D gauge coupling $g_{YM}$ and the little string scale,
\ie
\label{A'g}
{1\over 2\pi\A'} = {8\pi^2\over g_{YM}^2},
\fe
as seen by matching the tension of the instanton string with the fundamental string of DSLST, and also verified in \cite{Chang:2014jta}.  In this paper we work in units of $\A'$, and so $g_{YM}^2 = 32 \pi^3$.  Our convention for the Coulomb branch radius parameter $r$ is such that the $W$-boson corresponding to the D1-brane stretched between the $i$-th and $j$-th NS5 brane has mass
\ie
m_{ij} = 2r \left| \sin{\pi(i-j)\over k} \right|.
\fe
In the next two sections, we will study the expansion of the function $F_{a_1a_2a_3a_4}(s,t,u; r)$ in detail, from perturbative SYM and from DSLST.

\section{Perturbative 6D SYM in the Coulomb Phase}
\label{sec:sym}

Near the origin of the Coulomb branch, the $W$-bosons are light compared to the scale set by $g_{YM}$, and we can compute the 4-point amplitude of Cartan gluons in SYM perturbation theory. A priori, one may expect such a computation to run into two difficulties: the loop expansion of the massless scattering amplitude suffers from UV divergence at 4-loop order \cite{Bern:2012uf} (while the mixed Cartan gluon and $W$-boson amplitude diverges at 3-loop \cite{Bern:2010tq}), and there may be higher dimensional operators that deform the SYM Lagrangian \cite{Bossard:2010pk}. The consistency of DSLST \cite{Aharony:2003vk} combined with non-renormalization theorems of Sethi et al. implies that the SYM Lagrangian at the origin of the Coulomb branch is not deformed by ${\rm tr} F^4$ terms. The result of \cite{Chang:2014jta} further indicates that the $1/4$ BPS operator of the form $D^2 {\rm tr}^2 F^4$ is absent at the origin of the Coulomb branch as well. On the other hand, the 3-loop divergence in the non-Abelian sector means that the non-BPS dimension 10 operator $D^2 {\rm tr} F^4$ is needed as a counter-term \cite{Bern:2010tq}. Likewise, at 4-loop order we will need counter-terms of the form $D^4 {\rm tr} F^4$ and $D^4{\rm tr}^2 F^4$ \cite{Bern:2012uf}. It appears that one can proceed with the SYM perturbation theory, and add the appropriate counter-terms whenever a new divergence is encountered at a certain loop order. Of course, the perturbative SYM does not give a prescription for determining the finite part of these counter-terms. Such ambiguities however do not affect the leading logarithmic dependence on $r$, and so these leading logs can be computed unambiguously in the framework of SYM perturbation theory at small $r$. On the other hand, the finite shifts of the counter-terms that cannot be determined by SYM perturbation theory are in principle determined in the full little string theory, and one could hope for extracting such information from the opposite regime, namely the large $r$ limit.

Let us begin with the $F^4$ term in the Coulomb effective action, or more precisely, its supersymmetric completion, along the 1-dimensional subspace as specified in (\ref{zphi}). The corresponding superamplitude takes the form 
\ie
\delta^8(Q)  {C_{0,a_1a_2a_3a_4} \over r^2}.
\fe
As was shown in \cite{Aharony:2003vk}, the coefficient $C_0$ is given by 
\ie
C_{0,a_1a_2a_3a_4} = c_0\, {\rm min} \{ a_i, k-a_i\},~~~~0\leq a_i<k,
\fe
where $c_0$ is a constant that is independent of the color assignment.

Next consider the $D^2 F^4$ term, which can be written as
\ie
f_{1,a_1a_2a_3a_4}(r) s F_{a_1} \dotsb F_{a_4}
\fe
The result of \cite{Chang:2014jta} indicates that the corresponding superamplitude takes the form
\ie
\delta^8(Q) {1\over r^2} \Big[ C_{1,a_1a_2a_3a_4} s_{12} + (1\leftrightarrow 3) + (2\leftrightarrow 3) \Big],
\fe
and is {\it two-loop exact}.
By symmetry of permutation on external lines, $C_{1,a_1a_2a_3a_4}$ is invariant under the permutations $(12)$, $(34)$, as well as $(13)(24)$. Note that there is no 1-loop contribution to $f_1(r) D^2 F^4$, of order $r^{-4}$, simply because a 1-loop contribution would come with a $C_{1,a_1\cdots a_4}$ factor that is completely symmetric under permutation of $a_1,\cdots,a_4$, and thus must be proportional to $s+t+u$, which is zero.  Therefore $f_1(r)$ takes the simple form\footnote{When there is no potential confusion, we will often omit the color indices $a_1a_2a_3a_4$ if $a_1=a_2=-a_3=-a_4= \ell+1$. For example, $C_1 = C_{1,\, \ell+1, \, \ell+1 , \, -(\ell+1),\,-(\ell+1)}$.}
\ie
\label{f(r)D2F4}
f_1(r) = {C_1 \over r^2}.
\fe
In~\cite{Chang:2014jta}, the $C_1$ coefficients were computed for $k=2,3,4,5$ and color assignment $a_1 = a_2 = -a_3 = -a_4=\ell+1$ with $k-2\ge\ell\ge0$ .  The results are listed here in Table~\ref{tab:D2F4} with higher numerical precision.

\begin{table}[ht]
\begin{center}
\begin{tabular}{|c|c|c|}
\hline
$k$ & $\ell$ & $C_1$ \\\hline\hline
3 & 0, 1 & $-1.171954$ \\\hline
4 & 0, 2 & $-1.831931$ \\\hline
5 & 0, 3 & $-2.396790$ \\\hline
& 1, 2 & $-1.380352$ \\\hline
\end{tabular}
\end{center}
\caption{The coefficients in the small $r$ expansion of $f_1(r)$, which is the coefficient of $s F^4$.}
\label{tab:D2F4}
\end{table}

Now let us consider the $D^4 F^4$ term, which receives contributions from all loop orders. We can write the $D^4 F^4$ couplings as
\ie
f_{S,a_1a_2a_3a_4}(r) (s^2+t^2+u^2) F_{a_1}\cdots F_{a_4} + f_{A,a_1a_2a_3a_4}(r) s^2 F_{a_1}\cdots F_{a_4} 
\fe
where $S$ and $A$ stand for symmetric and asymmetric in the Mandelstam variables. $f_{S}(r)$ and $f_{A}(r)$ each admits a small $r$ expansion\footnote{
The color-ordered one-loop superamplitude is permutation invariant, hence the full amplitude is completely symmetric in $s$, $t$, and $u$, and so $C_{A}^1 = 0$.  The $\log^2$ divergence is also completely symmetric, as can be seen from (\ref{A4L}), and hence $B'_A = 0$.
}
\ie\label{nte}
& f_{S}(r) = {C_{S}^1\over r^6} + {C_{S}^2\over r^4} + {C_{S}^3\over r^2} + B_{S} \ln r + B'_{S} (\ln r)^2 + {\cal O}(r^2 (\ln r)),
\\
& f_{A}(r) = {C_{A}^2\over r^4} + {C_{A}^3\over r^2} + B_{A} \ln r + {\cal O}(r^2 (\ln r)).
\fe
The coefficients $C_{S/A}^1$, $C_{S/A}^2$, $C_{S/A}^3$ (which depend on the color factors) are computed from 1, 2, 3-loop amplitudes. The coefficients $B_{S/A}$ and $B'_{S/A}$ come from the 4-loop amplitudes, after canceling the log divergences by 3-loop and 4-loop counter-terms. Note that the appearance of the double log terms is due to nested divergences at 4-loop order. 
In the UV completed theory, namely the full LST, the divergence of SYM at 4-loop order and higher is reflected as a branch cut in the analytic structure of the function $f_2(r)$.
The detailed computation and numerical results for the 1, 2, and 3-loop contributions are given in Appendix~\ref{appA}, for $k=2,3,4,5$ and color assignment $a_1=a_2=-a_3=-a_4=\ell+1$ with $k-2\ge\ell\ge0$.  For 3-loop, we need to sum up the scalar integrals represented by the nine diagrams in Figure~\ref{fig:nine}.  Each of diagrams (e) (f) (g) (i) is in fact UV divergent by itself at linear order in $s$, $t$, $u$, and would potentially contribute to $D^2 F^4$.  However, these divergences cancel after we sum up these diagrams and the permutations of the external legs, and the remaining parts are quadratic or higher in $s$, $t$, $u$ and give finite contributions to $D^{2n} F^4$ for $n \geq 2$.

The 4-loop divergence can be computed at the origin of the Coulomb branch, as in \cite{Bern:2012uf}. After moving away from the origin on the Coulomb branch, in the expansion in external momenta, the logarithmic divergences appear in the form $\ln (\Lambda/r)$, and in the case of nested divergences, $(\ln(\Lambda/r))^2$. After canceling the logarithmic divergences with counter-terms, we are left with logarithmic dependence on $r$, and the coefficient of the leading log (or double log) is independent of finite shifts of the counter-term. 

The logarithmic divergence at the origin of the Coulomb branch involves three possible terms, of the form $(s^2+t^2+u^2){\rm tr} F^4$, $(s^2+t^2+u^2)({\rm tr} F^2)^2$, and $s^2 ({\rm tr} F^2)^2 + (2~{\rm more})$. The terms proportional to $(s^2+t^2+u^2)$ also contain double pole divergences (in dimensional regularization). To cancel the divergences we need a 3-loop counter-term $D^2 {\rm tr} F^4$ (it vanishes when restricted to the Cartan, but is now needed to cancel subdivergences in the 4-loop amplitude) and 4-loop counter-terms of the form $D^4 {\rm tr}F^4$ as well as $D^4({\rm tr} F^2)^2$. In the end, one obtain {\it unambiguously} the coefficient of
\ie
\ln r \left[s^2 ({\rm tr} F^2)^2+ (2~{\rm more})\right],
\fe
and the coefficient of 
\ie
(\ln r)^2 (s^2+t^2+u^2) ({\rm tr} F^2)^2 .
\fe
In principle, one can also determine unambiguously the $(\ln r)^2$ coefficient of the single trace term proportional to $(s^2+t^2+u^2) {\rm tr} F^4$, but this double pole coefficient has not been evaluated explicitly in \cite{Bern:2012uf}.

The $C^1_{S/A}$, $C^2_{S/A}$, $C^3_{S/A}$ and $B_{S/A}$ coefficients for $k = 2,3,4,5$ and color assignment $a_1 = a_2 = -a_3 = -a_4=\ell+1$ with $k-2\ge\ell\ge0$ are listed in Tables~\ref{tab:D4F4stu} and~\ref{tab:D4F4s}.

\begin{table}[ht]
\begin{center}
\begin{tabular}{|c|c|c|c|c|}
\hline
$k$ & $\ell$ & $C^1_{S}$ & $C^2_{S}$ & $C^3_{S}$ \\\hline\hline
2 & 0 & ${1 / 5760}$ & ${1 / 96}$ & 3.772838 \\\hline
3 & 0, 1 & $1 / 3240$ & ${1 / 36}$ & 7.086485 \\\hline
4 & 0, 2 &  $1 / 2304$ & ${3 / 64}$ & 11.619831 \\\hline
& 1 & $1 / 1440$ & $0.03590010$ & 8.521180 \\\hline
5 & 0, 3 & $1 / 1800$ & ${1 / 15}$ & 17.38894 \\\hline
& 1, 2 & $1 / 720$ & $0.06645686$ & 12.88988 \\\hline
\end{tabular}
\end{center}
\caption{The coefficients in the small $r$ expansion of $f_{S}(r)$, which is the coefficient of $(s^2+t^2+u^2) F^4$.}
\label{tab:D4F4stu}
\end{table}

~

\begin{table}[ht]
\begin{center}
\begin{tabular}{|c|c|c|c|c|}
\hline
$k$ & $\ell$ & $C^2_{A}$ & $C^3_{A}$ & $B_{A}$ \\\hline\hline
3 & 0, 1 & $-0.02459345$ & $-4.505248$ & 330.6754 \\\hline
4 & 0, 2 & $-0.04743323$ & $-8.729678$ & 541.8733 \\\hline
5 & 0, 3 & $-0.06993323$ & $-13.955903$ & 839.61925 \\\hline
& 1, 2 & $-0.04593824$ & $-8.901921$ & 419.8096 \\\hline
\end{tabular}
\end{center}
\caption{The coefficients in the small $r$ expansion of $f_{A}(r)$, which is the coefficient of $s^2 F^4$.}
\label{tab:D4F4s}
\end{table}

\section{The $\A'$ Expansion of Little String Amplitude}

The vertex operators of the massless Cartan gluons in double scaled little string theory are in the (R,R) sector, of the form \cite{Aharony:2003vk,Chang:2014jta},
\ie
{\cal V}_{a\dot b,\ell}^\pm = e^{-{\varphi\over2}-{\widetilde\varphi\over 2}} e^{ip_\mu X^\mu} \lambda^A_a\widetilde\lambda_{B\dot b} S_A \widetilde S^B V^{sl,(\mp{1\over 2},\mp{1\over 2})}_{{\ell\over 2},\pm {\ell+2\over 2},\pm {\ell+2\over 2}} V^{su,(\pm{1\over 2},\pm{1\over 2})}_{{\ell\over 2},\pm {\ell\over 2},\pm {\ell\over 2}},
\fe
with $\ell = 0,1,\cdots,k-2$ labeling the color index of the $U(1)^{k-1}$ gluons according to their eigenvalues $e^{2\pi i(\ell+1)/k}$ with respect to the $\mathbb{Z}_k$ cyclic permutation of the NS5 branes. $\lambda^A_a$ and $\widetilde\lambda_{B\dot b}$ are the 6D spinor helicity variables as before, and $S_A,\widetilde S^B$ are the left and right spin fields of the $\mathbb{R}^{1,5}$ part of the worldsheet CFT. There is also an identification ${\cal V}^-_{a\dot b,\ell} \equiv {\cal V}^+_{a\dot b,k-2-\ell}$ \cite{Fateev:1985mm,Aharony:2003vk,Chang:2014jta}. It was shown in \cite{Chang:2014jta} that the sphere 4-point superamplitude takes the form
\ie
\label{ALST}
& {\cal A}_{DSLST}(1^{\ell+1}, 2^{\ell+1}, 3^{-\ell-1},4^{-\ell-1}) = \delta^8(Q) \,{\cal N}_{k,\ell} \int_{\mathbb C} d^2z |z|^{{(\ell+1)^2\over k} - s - {1\over 2}} |1-z|^{\ell - {(\ell+1)^2\over k} - u+{1\over 2} }
\\
&~~~~~~\times \left\langle V^{su,({1\over 2},{1\over 2})}_{{\ell\over 2}, {\ell\over 2}, {\ell\over 2}} (z,\bar z) V^{su,({1\over 2},{1\over 2})}_{{\ell\over 2}, {\ell\over 2}, {\ell\over 2}} (0)
V^{su,(-{1\over 2},-{1\over 2})}_{{\ell\over 2}, -{\ell\over 2}, -{\ell\over 2}} (1)
V^{su,(-{1\over 2},-{1\over 2})}_{{\ell\over 2}, -{\ell\over 2}, -{\ell\over 2}} (\infty)\right\rangle_{SU(2)_k/U(1)}
\\
&~~~~~~\times \int_0^\infty {dP\over 2\pi} C(\A_1,\A_2, {Q\over 2}+iP) C(\A_3,\A_4, {Q\over 2}-iP) 
\left| F(\Delta_1,\Delta_2,\Delta_3,\Delta_4;\Delta_P;z) \right|^2.
\fe
Here ${\cal N}_{k,\ell}$ is a normalization constant, $C(\A_1,\A_2,\A_3)$ is the structure constant of Liouville primaries, and $F(\Delta_1,\Delta_2,\Delta_3,\Delta_4;\Delta_P;z)$ is the Liouville 4-point conformal block. See \cite{Chang:2014jta} for the precise identification of the parameters $\A_i$, $\Delta_i$ etc.

The evaluation of the conformal block integral and the integration over the cross ratio $z$ are performed numerically, order by order in the $\A'$ expansion.\footnote{
Since we set $\A' = 1$, the $\A'$ expansion is an expansion in the Mandelstam variables. 
}
For $k=2,3,4,5$, the two leading terms in the expansion were given in \cite{Chang:2014jta}. We carry out this computation to $\A'^3$ order, with the order $\A'^n$ terms corresponding to $D^{2n} F^4$ coupling in the Coulomb branch effective action. In the following we normalize the amplitudes by their $\A'^0$ order terms.

\begin{itemize}
\item  $\mathbf{k = 2, ~\ell = 0}:$
\ie
1 + 2.10359958 (s^2+t^2+u^2) + 17.42982502 stu +\cdots.
\fe
\item  $\mathbf{k = 3, ~\ell = 0}:$
\ie
1 -1.171954 s + 5.20891 (s^2+t^2+u^2) - 4.88324 s^2 + 63.814 stu -20.8624 s^3+\cdots.
\fe
\item  $\mathbf{k = 4, ~\ell = 0,2}:$
\ie
& 1 - 1.83193119 s + 9.466198 (s^2+t^2+u^2) - 9.334781 s^2 
\\
& \hspace{2.5in} + 153.967791 stu -51.209842 s^3+\cdots.
\fe
\item  $\mathbf{k = 4, ~\ell = 1}:$
\ie
1 + 6.1080323 (s^2+t^2+u^2) + 96.795814 stu+\cdots.
\fe
\item  $\mathbf{k = 5, ~\ell = 0,3}:$
\ie
1 - 2.39679 s + 14.9055 (s^2+t^2+u^2) - 14.8295 s^2 + 302.54 stu -100.798 s^3+\cdots.
\fe
\item  $\mathbf{k = 5, ~\ell = 1,2}:$
\ie
1 - 1.38035 s + 10.3118 (s^2 + t^2 + u^2) - 9.4101 s^2 + 202.166 stu - 65.509 s^3+\cdots.
\fe
\end{itemize}
The omitted terms are of quartic and higher degrees in $s,t,u$, corresponding to $D^8 F^4$ and higher derivative couplings in the effective action.

Note that the DSLST four-point amplitude is invariant under flipping the $\mathbb Z_k$ charges of the vertex operators. In addition when $\ell+1=k/2$ (i.e. the vertex operators are identical), the amplitude is invariant under permutation of the Mandelstam variables.

\subsection{An interpolating function from weak to strong coupling}

On one hand, perturbative 6D SYM gives a small $r$ expansion of the coefficient $f_2(r)$ of each higher derivative term in the Coulomb branch effective action. 
On the other hand, perturbative string scattering in DSLST gives an expansion valid at large $r$.
The exact $f_2(r)$ is a function that interpolates the two ends.

Let us first consider the $D^2 F^4$ term.  A non-renormalization theorem by~\cite{Sethi:1999qv} shows that $f_1(r)$ is two-loop exact in $SU(3)$ maximal SYM, which means that (\ref{f(r)D2F4}) should hold for arbitrary $r$.  Indeed, the result of \cite{Chang:2014jta} was that the coefficients of $s$ in the tree-level DSLST superamplitudes exactly match with the $C_1$ obtained from the SYM two-loop superamplitudes (see Table~\ref{tab:D2F4}), for $k=4, 5$ as well as $k=3$.  It is not inconceivable that $f_1(r)$ is two-loop exact in 6D SYM for all $k$, which also implies that all higher genus superamplitudes for the scattering of four Cartan gluons should vanish at ${\A'}^1$ order.

\begin{figure}[t!]
\begin{center}
\includegraphics[width=.45\textwidth]{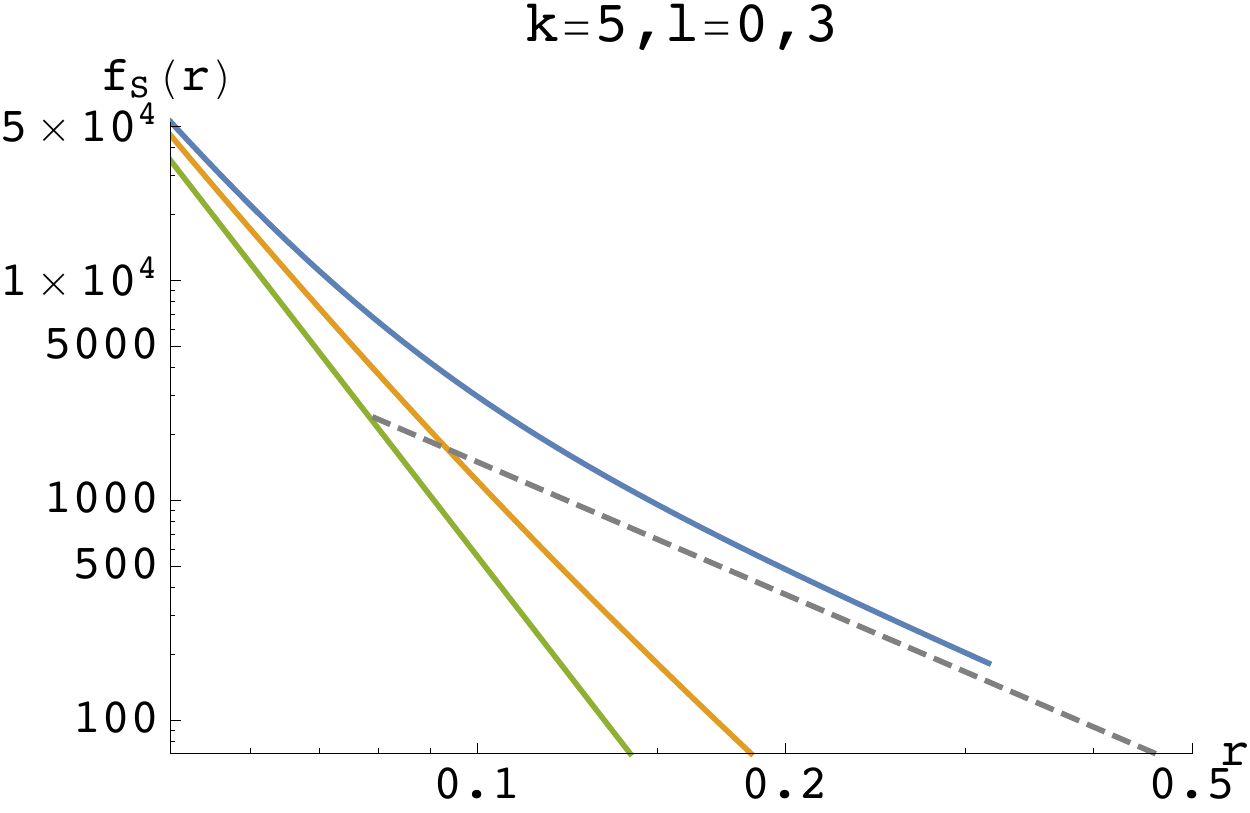}
\includegraphics[width=.45\textwidth]{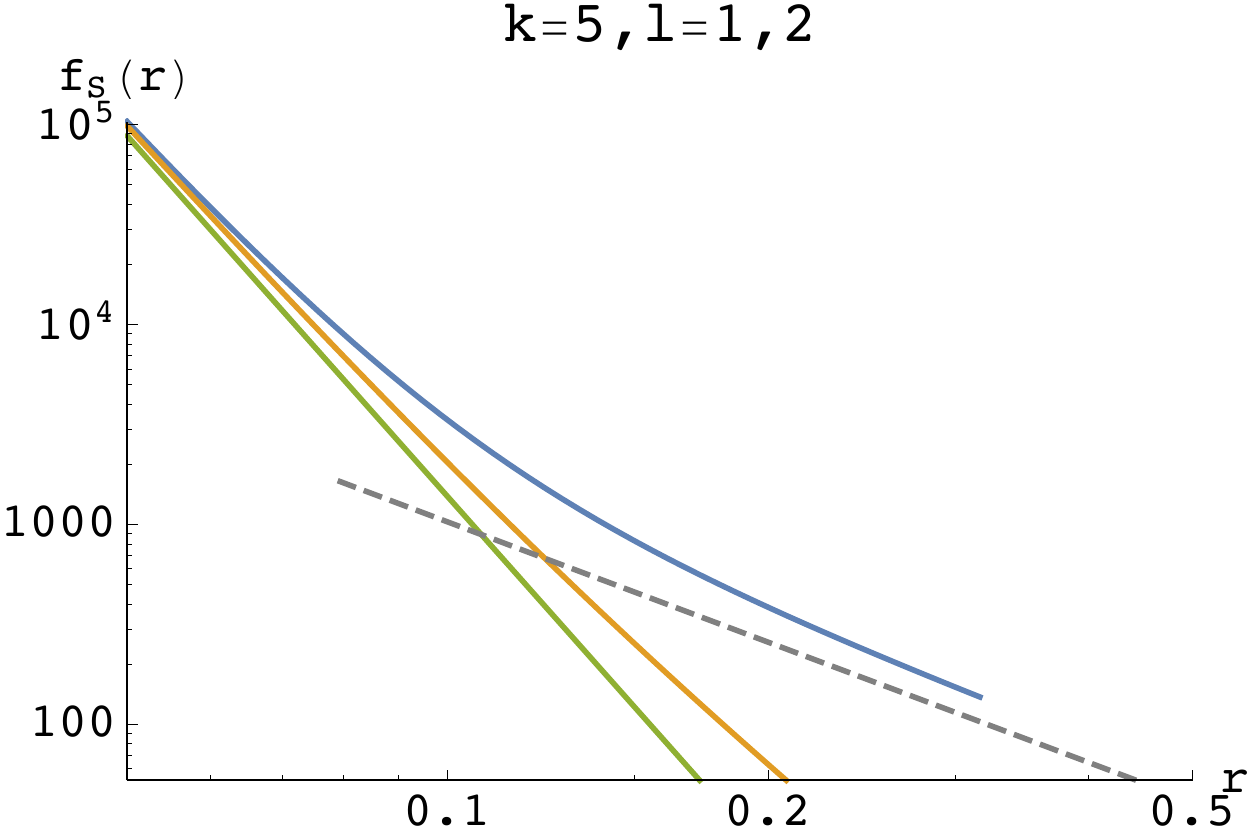}
\end{center}
\vspace{.1in}
\begin{center}
\includegraphics[width=.45\textwidth]{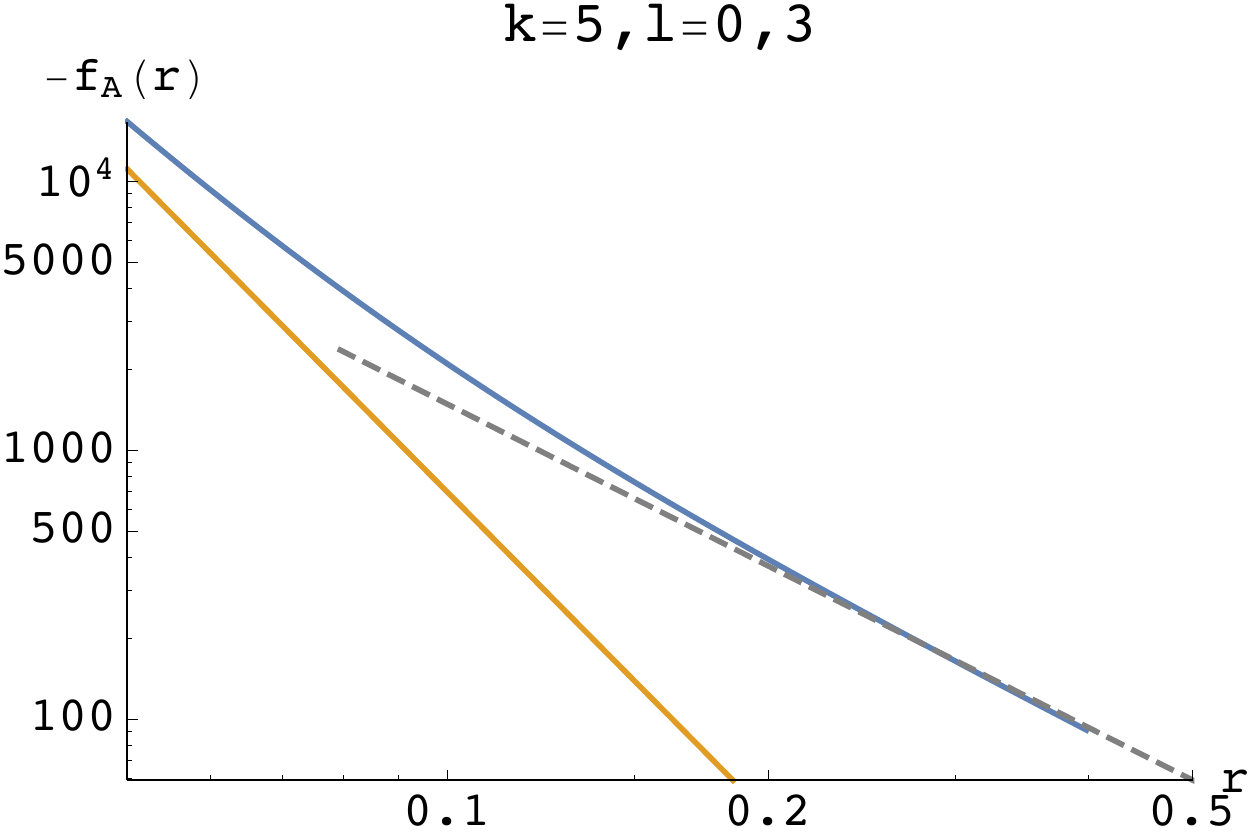}
\includegraphics[width=.45\textwidth]{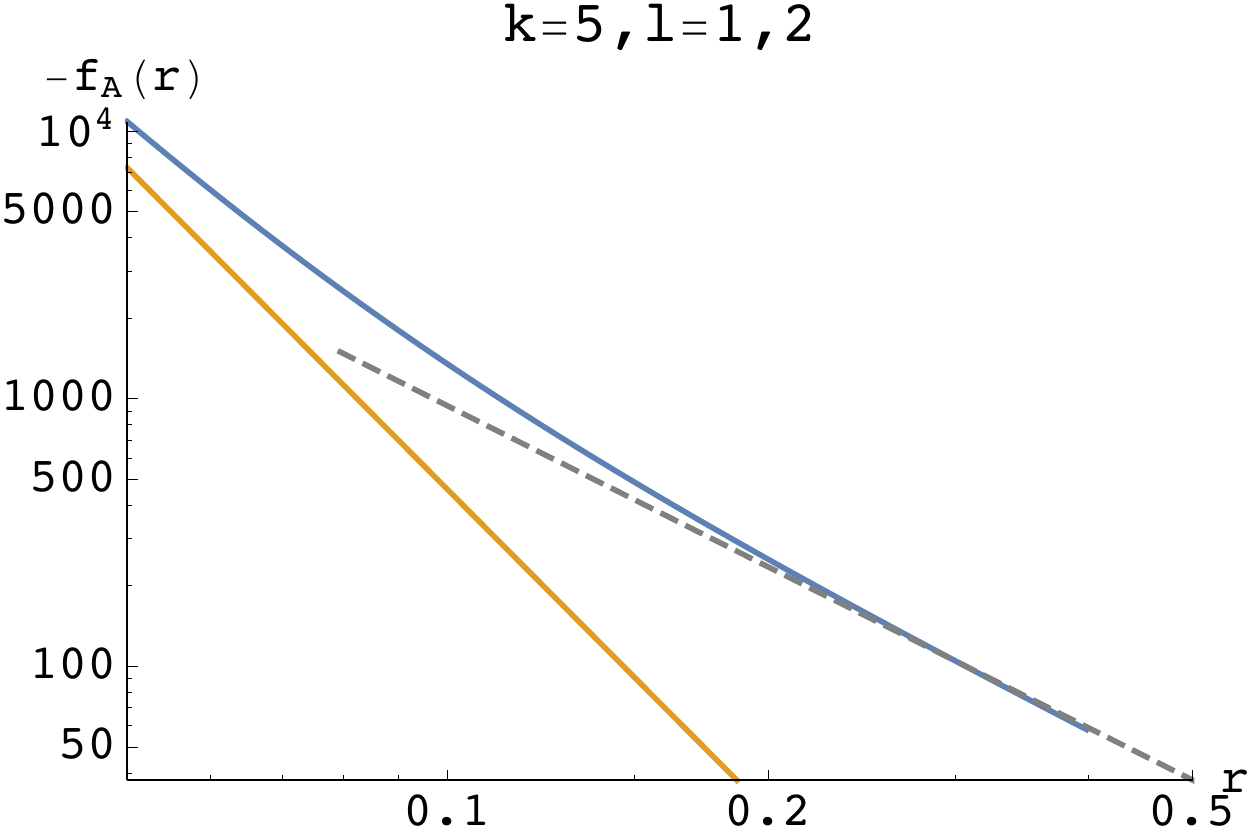}
\end{center}
\caption{Log-log plot of the coefficient $f_S(r)$ of $(s^2+t^2+u^2)F^4$ and $f_A(r)$ of $s^2 F^4$. The dashed line is given by the DSLST tree level superamplitude (valid for large $r$). The lower green line comes from 6D SYM one loop, the middle orange line comes from one and two loops combined, and the upper blue line combines the contributions up to three loops (valid for small $r$).  We interpolate the two ends by a naive extension beyond their regimes of validity.
}
\label{fig:interp}
\end{figure}


Next let us consider $D^4 F^4$. With the color index assignment $a_1=a_2=-a_3=-a_4=\ell+1$ (labeling the $\mathbb{Z}_k$ charge), the two independent structures are proportional to $s^2+t^2+u^2$ and $s^2$. We will compare the large and small $r$ expansions. 
On the 6D SYM side, the $r^a (\ln (r/\Lambda))^b$ terms after resummation will correct the power of $r$ when one interpolates the function $f_2(r)$ to large $r$. Here the coefficient of $(\ln(r/\Lambda))^2$ in the small $r$ expansion can be determined by the 4-loop UV divergence at the origin of the Coulomb branch. However, as already mentioned, this computation involves the divergence in the single trace $D^4 {\rm tr} F^4$ term, which has not yet been computed in 6D SYM. The scale $\Lambda$ has absorbed the contribution from the counter term, and is expected to be of order $g_{YM}^{-1}$ in the full LST. Since the actual numerics depends on the precise value of the mass scale $\Lambda$, we will not include the $\ln (r/\Lambda)$ terms in the interpolation function.

In each case, the coefficients of $1/r^2$ are close but not equal between the large and small $r$ expansions. There is no reason for them to be equal, since the large $r$ expansion should be corrected by higher genus contributions of order $1/r^{2(g+1)}$, and the small $r$ expansion includes one-loop $1/r^6$ and two-loop $1/r^4$ terms, and should further be corrected by higher-loop contributions of the form $r^a (\log (r/\Lambda))^b$.  


For concreteness, we explicitly make the comparison for $k = 5$, noting that the other cases are qualitatively the same.
\begin{itemize}
%
%
\item  $\bf k = 5, ~ \ell = 0:$
Large $r$ expansion:
\ie
f_S(r) &= {14.9055 \over r^2} + {\cal O}(1/r^4),
\\
f_A(r) &= - {14.8295 \over r^2} + {\cal O}(1/r^4) .
\fe
Small $r$ expansion:
\ie
f_S(r) &= {1 \over 1800 r^6} + {1 \over 15 r^4} + {17.38894 \over r^2} + {\cal O}(\ln (r/\Lambda),(\ln (r/\Lambda))^2),
\\
f_A(r) &= - {0.06993323 \over r^4} - {13.955903 s^2 \over r^2} + 839.61925 \ln (r/\Lambda)+ {\cal O}(r^2).
\fe
\item  $\bf k = 5, ~ \ell = 1:$
Large $r$ expansion:
\ie
f_S(r) &= {10.3118 \over r^2} + {\cal O}(1/r^4),
\\
f_A(r) &= -{9.4101 \over r^2} + {\cal O}(1/r^4).
\fe
Small $r$ expansion:
\ie
f_S(r) &= {1 \over 720 r^6} + {0.06645686 \over r^4} + {12.88988 \over r^2} + {\cal O}(\ln (r/\Lambda),(\ln (r/\Lambda))^2),
\\
f_A(r) &= - {0.04593824 \over r^4} - {8.901921 \over r^2} + 419.8096 \ln (r/\Lambda) + {\cal O}(r^2).
\fe
\end{itemize}
In Figure~\ref{fig:interp}
, the large $r$ expression is plotted in dashed lines, and the small $r$ (up to 1, 2, and 3 loops) are plotted in solid lines.  We interpolate the two ends by a naive extension beyond their regimes of validity.


\section{Discussion}

To summarize our results so far, while the $r^{-2} F^4$ and $r^{-2} D^2 F^4$ terms in the Coulomb branch effective action are computed {\it exactly} by perturbative SYM at one-loop and two-loop orders respectively, and match precisely with the corresponding $\A'$-expansion of the tree level amplitude in DSLST, the $f_2(r) D^4 F^4$ terms involve a set of nontrivial interpolation functions $f_2(r)$, that receive a priori all-loop contribution in SYM perturbation theory. We have determined $f_2(r)$ in its small $r$ expansion up to 3-loop orders in 6D SYM. Interestingly, the 3-loop contribution that scales like $r^{-2}$, is numerically close (but not equal) to the result obtained from $\A'^2$ order terms in the tree amplitude of DSLST, which captures the large $r$ limit of $f_2(r)$.

Starting at 4-loop order in the perturbative SYM description, one encounters UV divergences and while the leading log coefficients can be determined unambiguously in perturbation theory, the subleading logs and constant shifts depend on finite parts of 3 and 4-loop counter terms ($D^2 {\rm tr} F^4$, $D^4 {\rm tr} F^4$, and $D^4{\rm tr}^2 F^4$ at the origin of the Coulomb branch), and are a priori {\it undetermined} in 6D SYM perturbation theory.

In principle, the $(1,1)$ LST provides an unambiguous UV completion of the perturbative amplitudes of 6D SYM. If one could somehow compute the exact 4-gluon amplitude in DSLST, non-perturbatively in $g_s$, then one should recover all the perturbative SYM loop amplitudes, and fix the finite parts of all counter terms. While we do not have the technology for such exact computations on the string theory side, the interpolation results on the Coulomb moduli space so far suggests that, despite the non-renormalizability of the 6D SYM, the naive perturbative expansion is a valid prescription provided that appropriate counter terms are included at each loop order.\footnote{For instance, one could have said that since the 6D SYM theory is expected to be strongly coupled at the scale $g_{YM}^{-1}$, a UV cutoff should be imposed at the scale $\Lambda\sim g_{YM}^{-1}$, and there would seem to be no reason to perform the loop integral over momenta above this scale. However, the exact agreement of one-loop and two-loop contributions to the $F^4$ and $D^2 F^4$ terms with DSLST indicates that the naive loop integrals, which happen to be free of UV divergences in these cases, give the correct answer.}

The UV divergences that arise at 4-loop order and higher in the massless amplitudes of 6D SYM in the Coulomb phase, indicate not a trouble with SYM perturbation theory, but rather a {\it feature} of the amplitudes and the corresponding couplings in the Coulomb branch effective action. Namely, the function $f_2(r)$, as an analytic function of $r$ on the Coulomb branch, has a branch cut starting from the origin. Where does this branch cut end, in the analytic continuation of Coulomb moduli space? A natural expectation is that perhaps the branch cut goes all the way to $r=\infty$, where the Coulomb phase is described by weakly coupled DSLST. In fact, we generally expect non-analyticity in $f_2(r)$ at $r=\infty$, due to the non-convergence of the string perturbation series, and the need for stringy non-perturbative contributions (e.g. D-instanton amplitudes). In fact, due to the identification $g_s\sim 1/r$, we could speculate that non-perturbative string amplitudes of the form $\exp(-1/g_s)\sim e^{-r}$, contributes to the finite counter terms at the origin of the Coulomb moduli space!

Going beyond massless amplitudes, the scattering of gluons with $W$-bosons in 6D SYM may be compared to D-brane scattering amplitudes in DSLST.\footnote{At the level of 3-point amplitude of gluon emission by a $W$-boson, the agreement with the disc 1-point amplitude in DSLST was known in \cite{Israel:2005fn}.} We hope to report on these results in the near future.

\section{Comments on $(2,0)$ LST and 5D SYM}

In this section we discuss the compactification of the (2,0) DSLST to five dimensions, and constrain the higher derivative terms in the effective action of the resulting 5D gauge theory on the Coulomb branch. In particular, we will show that the $\text{tr}F^4$ coupling at the origin of the Coulomb branch of the circle-compactified (2,0) superconformal field theory is absent.

At the perturbative level, or equivalently in the $1/r$ expansion on the Coulomb branch, the structure of $(2,0)$ DSLST is very similar to $(1,1)$ DSLST, differing only through GSO projection. As far as the massless 4-point amplitude is concerned, at string tree level the only difference between the $(2,0)$ and $(1,1)$ case is the interpretation of the supermomentum delta function $\delta^8(Q)$ in terms of the polarizations of the massless supermultiplets involved. The scalar function of $s,t,u$ that multiplies $\delta^8(Q)$ is identical. An analogous statement holds for the genus one 4-point amplitude as well. In the NSR formalism, this can be seen by noting that the contribution from the (P,P) spin structure vanishes,\footnote{
It suffices to look at the scattering of the scalars which correspond to (NS,NS) vertex operators. At one loop in the (P,P) sector (here we are following the convention of \cite{Polchinski:1998rr} although historically this had been also referred as the (odd,odd) sector \cite{D'Hoker:1988ta}), we need to have three $(0,0)$-picture and one $(-1,-1)$-picture vertex operators plus one PCO. Hence in the path integral we have a total of $4$ insertions of $\psi^\m$ and $\tilde\psi^\m$ which leads to a vanishing contribution to the total amplitude due to the presence of six zero modes for $\psi^\m$ (and $\tilde{\psi}^\m$).

One can also reach the slightly stronger statement that the four point amplitudes in $(1,1)$ and $(2,0)$ DSLST agree up to 2-loops following a version of Berkovits' argument in Section 3.2 in \cite{Berkovits:2006vc}.
}
and therefore the IIA and IIB GSO projections yield the same amplitudes, up to reassignment of polarization tensor structure. It is not inconceivable that the massless 4-point amplitudes in $(2,0)$ and $(1,1)$ DSLST involve the same scalar function of $s,t,u$ to all order in perturbation theory, though we do not have an argument for this. On the other hand, it appears that the D-instanton amplitudes of massless string scattering will be quite different in the two theories, as the BPS D-instanton that is pointlike in the $\mathbb{R}^6$ and localized at the tip of the cigar exists only in the $(2,0)$ DSLST and not in the $(1,1)$ theory. Such contributions could alter the effective action near the origin of the Coulomb branch significantly, and give rise to entirely different low energy dynamics of the $(2,0)$ and $(1,1)$ LST at the origin of the Coulomb branch.

Nonetheless, in view of the idea that the $(2,0)$ SCFT, when compactified on a circle, is described in the low energy limit as 5D maximally supersymmetric Yang-Mills theory \cite{Seiberg:1997ax} together with an infinite series of higher dimensional operators/counter-terms \cite{Lee:2000kc,Lambert:2010iw,Douglas:2010iu,Lambert:2012qy}, one could ask whether there is a similar interpolation on the Coulomb branch of the $(2,0)$ LST compactified on a circle. 
 In this case, the $W$-boson  comes from D-branes located at the tip of the cigar in the T-dual picture \cite{Israel:2005fn}. The parameters in the circle-compactified DSLST are the string length $\ell_s$, the $W$-boson mass $m_W$ which is related to the string coupling\footnote{In this paper, we use $g_s$ to denote the string coupling at the tip of the cigar in the IIB picture , not to be confused with the asymptotic string coupling $g_s^\infty$ before taking the decoupling limit of NS5-branes in asymptotically flat spacetime in the IIA picture. They are related by $g_s\sim { \ell_s g_s^{\infty}/ r }$ \cite{Giveon:1999px,Ooguri:1995wj,Kutasov:1995te,Sfetsos:1998xd}
.
}  $g_s$ by
\ie\label{mwr}
m_W \sim {R\over g_s\ell_s^2},
\fe
and the compactification radius $R$, which is related to the 5D gauge coupling $g_5$ by
\ie
R = {g_5^2\over 8\pi^2}.
\fe
From the 5D perspective, the natural mass scale is set by $g_5$ or $R$, and the two dimensionless parameters are $\rho \sim m_W R$ (parameterizing distance from the origin on the Coulomb branch) and $R/\ell_s$. The 5D gauge theory obtained from compactification of $(2,0)$ SCFT, in its Coulomb phase, is obtained in the limit $R/\ell_s\to \infty$, while holding $R$ and $\rho$ fixed.\footnote{Note that it is a different limit than taking $R/\ell_s\rightarrow \infty$ while keeping $g_s$ and $\ell_s$ fixed, which is the limit of decompactified (2,0) DSLST.}
 This in particular requires sending $g_s\to \infty$ at the same time.

If we write the amplitude of massless particles in the compactified $(2,0)$ DSLST in the form
\ie
{\cal A}_{(2,0) \,{\rm DSLST}}(g_s, E^2\ell_s^2, ER),
\fe
and the corresponding amplitude in the UV completion of 5D SYM in the form
\ie
{\cal A}_{\rm 5D\, GT}(g_5^2 m_W, g_5^2 E),
\fe
then we expect
\ie\label{syc}
\lim_{g_s\to\infty} {\cal A}_{(2,0) \,{\rm DSLST}}\left(g_s, {E^2 g_5^2\over g_s m_W}, g_5^2 E\right) = {\cal A}_{\rm 5D\, GT}(g_5^2 m_W, g_5^2 E).
\fe
The LHS cannot be captured by DSLST perturbation theory in a straightforward manner. 
For instance, we can write the $D^{2n} F^4$ terms in the Coulomb branch quantum effective Lagrangian in the schematic form
\ie
\sum_{n=0}^\infty f_n(\rho) D^{2n}F^4 ,
\fe
where $\rho \sim m_W R$ is the distance parameter on the Coulomb branch, and the subscript $n$ indicates the ``number of derivatives". If we assume that the UV completion of the 5D SYM perturbation theory is such that higher dimensional operators are added only when needed as counter-terms,\footnote{As we will see shortly, while this is expected for the compactified $(2,0)$ SCFT, this is not true for the compactified $(2,0)$ LST. We thank C. C\'{o}rdova and T. Dumistrescu for a key discussion on this point.} then the SYM loop expansion of $f_n(\rho)$ has the structure
\ie\label{fff}
& f_0(\rho) = {f_0^{(1)}\over \rho^3},
\\
& f_1(\rho) = {f_1^{(2)}\over \rho^4} + {f_1^{(3)}\over \rho^3}+\cdots ,
\\
& f_2(\rho) = {f_2^{(1)}\over \rho^7} + {f_2^{(2)}\over \rho^6} + {f_2^{(3)}\over \rho^5} + \cdots + {f_2^{(8)}}\ln \rho + \cdots.
\fe
Here the coefficient $f_n^{(L)}$ comes from the $L$-loop 4-point amplitude. Note that the 1-loop contribution $f_1^{(1)}/\rho^5$ is absent; this is because the 1-loop amplitude involves only a single color structure that is invariant under permuting the 4 external lines, and the $D^2 F^4$ amplitude would be proportional to $s+t+u$ which vanishes. Note that while the 5D SYM 4-point amplitude is known to have UV divergence at 6-loop order \cite{Bern:2012di}, such a divergence vanishes when the external gluons are restricted to the Cartan subalgebra. This is because the counter-term responsible for this divergence is the unique dimension 10 non-BPS operator of the form $D^2 {\rm tr} F^4+\cdots$ \cite{Bossard:2010pk,Chang:2014kma}, which in fact vanishes upon Abelianization (i.e. restricting to the Cartan subalgebra). The 4-point amplitude of Cartan gluons in 5D SYM is expected to diverge first at 8-loop order, with the counter-term being a non-BPS operator of the form $D^4 \tr F^4+\cdots$. In the UV completion that is expected to arise from the compactification of $(2,0)$ theory, the $D^4\tr F^4$ counter-term should cancel the log divergence, leaving a $\ln \rho$ dependence in the Coulomb effective action, hence the $f_2^{(8)}\ln \rho$ term in (\ref{fff}).

Let us focus on the $f_0(\rho) F^4$ coupling for the moment. The argument of \cite{Paban:1998qy} and \cite{Maxfield:2012aw} indicates that, at least in the $SU(2)$ case where the Coulomb branch moduli space is just a single $\mathbb{R}^5$, $f_0(\rho)$ is a harmonic function on the $\mathbb R^5$.\footnote{This is consistent with the $v^4/ \rho^3$ effective potential between separate D4 branes moving at a relative velocity \cite{Douglas:1996yp}.} Assuming $SO(5)$ R-symmetry, such a harmonic function must be of the form
\ie
f_0(\rho) = c + {f_0^{(1)}\over \rho^3}.
\fe
The constant $c$, if non-vanishing, would correspond to a ${\rm tr} F^4$ coupling in the non-Abelian SYM at the origin of the Coulomb branch moduli space. In writing (\ref{fff}) we have assumed that such coupling is absent in the low energy limit of the compactified $(2,0)$ theory. We will now justify this assumption. 

The Coulomb phase of the circle-compactified $A_1$ $(2,0)$ LST has a moduli space of vacua $\mathbb{R}^4\times S^1$. The $S^1$ coming from the compact scalar in the 6D $(2,0)$ tensor multiplet, and has size $\sim (R/\ell_s)^2$ in units of $R$.\footnote{To see the size of the $S^1$, we can go back to the NS5-brane picture in type IIA string theory, separated in the transverse $\mathbb{R}^4$, with the world volume of the NS5-branes compactified on a circle of radius $R$. A $W$-boson coming from D2-brane stretched between a pair of the NS5-branes and wrapping the circle has mass $m_W \sim R r /(g_s^\infty \ell_s^3)\sim R/(g_s \ell_s^2)$ as before. On the other hand, if we are to separate the NS5-branes along the M-theory circle, the M2-brane stretched between the M5-branes and wrapping the compactification circle of radius $R$ has mass $\sim R/\ell_s^2$.}
In the Coulomb phase of the compactified $(2,0)$ LST, the $D^{2n} F^4$ couplings come with the coefficients $f_{n}(\vec \rho, R/\ell_s)$, such that
\ie
\lim_{R/\ell_s \to \infty} f_{n}(\vec \rho,R/\ell_s) = f_{n}(\rho).
\fe
Here $\vec \rho$ parameterize a point on the $\mathbb{R}^4\times S^1$ moduli space, and the function $f_{n}(\vec \rho, R/\ell_s)$ is invariant under $SO(4)$ $R$-symmetry in 6 dimensions, while the $SO(5)$ is only restored in the $R/\ell_s\to \infty$ limit. Note that, importantly, the limit is taken with $\rho=R^2/(g_s\ell_s^2)$ held fixed, and so taking $R/\ell_s\to \infty$ requires sending $g_s\to \infty$ at the same time. From the 5D perspective, $g_s$ of DSLST is related to the vev of a massless scalar field, whereas $R/\ell_s$ is a rigid parameter (there is no massless graviton propagating in the $\mathbb{R}^{1,5}$ of the DSLST and hence there is no massless 5D scalar associated with the compactification radius); in particular, the dependence on $g_s$ is constrained by supersymmetry, whereas the dependence on $R/\ell_s$ is not.

At finite $R/\ell_s$, $f_0(\vec \rho,R/\ell_s)$ is an $SO(4)$-invariant harmonic function on the $\mathbb{R}^4\times S^1$. We can write $\vec \rho=(\vec \Phi, y)$, where $\vec \Phi$ parameterizes the $\mathbb{R}^4$ and $y$ is the coordinate on the $S^1$. The harmonic function $f_0(\vec \rho,R/\ell_s)$ is restricted to be of the form
\ie
f_0(\vec \rho,R/\ell_s) = c(R/\ell_s) + \sum_{n\in\mathbb{Z}} {f_0^{(1)}(R/\ell_s) \over \left[ |\Phi|^2 + \left(y+ n {R^2\over \ell_s^2} \right)^2 \right]^{{3\over 2}}}.
\fe
While $c$ may no longer be a constant, it must be a function of the rigid parameter $R/\ell_s$ only.
In the limit of large $|\Phi|$, $f_0$ can be expanded as
\ie
f_0(\vec \rho,R/\ell_s) = c(R/\ell_s) +  {2\ell_s^2 f_0^{(1)} (R/\ell_s) \over R^2}{1\over |\Phi|^2} +\cdots.
\fe
Matching this with the tree level $(2,0)$ DSLST, we conclude that $c(R/\ell_s)\equiv 0$. From this argument we also expect that the corrections to the tree level contribution to $F^4$ coupling in the compactified DSLST are entirely non-perturbative in $g_s$.

Now, near the origin of Coulomb branch, $(\Phi,y)=(0,0)$, $f_0$ can be written as
\ie
f_0(\vec \rho,R/\ell_s) &= f_0^{(1)}(R/\ell_s) \left\{ {1\over \rho^3} +  \sum_{n\not=0} { \left[ |\Phi|^2 + \left(y+ n {R^2\over \ell_s^2} \right)^2 \right]^{-{3\over 2}}} \right\}
\\
&= f_0^{(1)}(R/\ell_s) \left\{ {1\over \rho^3} + 2\zeta(3) \left({\ell_s\over R}\right)^6 + 3 \zeta(5) \left( 4y^2- |\Phi|^2 \right) \left({\ell_s\over R}\right)^{10} + \cdots \right\}.
\fe
The first term proportional to $\rho^{-3}$ is generated from 5D SYM by integrating out $W$-bosons at 1-loop. The second term is non-vanishing at the origin of the Coulomb branch and can be understood in terms of 6D SYM compactified on a circle (as in the T-dual $(1,1)$ LST), with massive Kaluza-Klein modes integrated out at 1-loop. This term vanishes in the $R/\ell_s\to \infty$ limit, and thus the ${\rm tr} F^4$ coupling is absent in the compactified $(2,0)$ superconformal theory (at the origin of its Coulomb branch). The third term comes from the 1-loop diagram with 6D $W$-bosons in the loop that also carry nonzero KK momenta, expanded to the second order in the $W$-boson mass parameter, and gives rise to an $SO(5)_R$ breaking dimension 10 BPS operator at the origin of the Coulomb branch of the 5D gauge theory.

It should be possible to extend this discussion to higher rank cases as well. A more detailed investigation of the two-parameter interpolation function in the Coulomb phase of compactified $(2,0)$ DSLST, and its interplay with the perturbative structure of 5D SYM, are left to future work.

\bigskip

\section*{Acknowledgments}
We would like to thank Chi-Ming Chang for collaboration during the early stage of this project. We are grateful to Ofer Aharony, Lance Dixon, Daniel Jafferis, Cumrun Vafa for useful conversations and correspondences, to Zohar Komargodski and Shiraz Minwalla for key suggestions on the use of non-renormalization theorems in the Coulomb effective action, to Clay C\'{o}rdova and Thomas Dumitrescu for important discussions on the compactification of $(2,0)$ theories, and to Travis Maxfield and Savdeep Sethi for comments on a preliminary draft. We would like to thank the 7th Taiwan String Workshop at National Taiwan University, Weizmann institute, and Kavli IPMU for their support during the course of this work. The numerical evaluations of loop integrals are performed using FIESTA on the Harvard Odyssey cluster, whereas the conformal block integrations and DSLST amplitudes are computed with Mathematica. S.H.S. is supported by the Kao Fellowship at Harvard University.  X.Y. is supported by a Sloan Fellowship and a Simons Investigator Award from the Simons Foundation. Y.W. is supported in part by the U.S. Department of Energy under grant Contract Number  DE-SC00012567.  This work is also supported by NSF Award PHY-0847457, and by the Fundamental Laws Initiative Fund at Harvard University.

\appendix

\section{6D SYM Loop Amplitudes Contributing to $D^4 F^4$}
\label{appA}

The term $f_2(r) D^4F^4$ receives contribution from all loop orders of the scattering amplitude of four Carton gluons.  At each loop order, we need expand the superamplitude to quadratic order in the Mandelstam variables.  Each loop order is proportional to the {\it color-ordered} four-point tree-level scattering amplitude
\begin{align}
\mathcal{A}^{tree}(1,2,3,4) = - {i\over s_{12}s_{14} } \delta^8 (Q).
\end{align}

\subsection{One-loop}

\begin{figure}[htb]
\begin{center}
\includegraphics[scale=1.1]{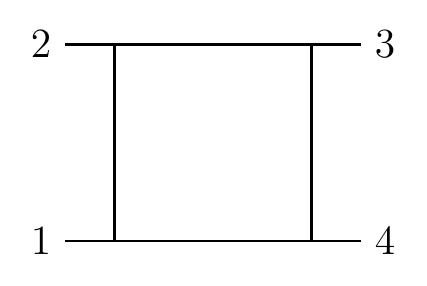}
\end{center}
\caption{The 1-loop scalar integral $I_4^{1-loop}(s_{12},s_{14},m_{ij})$.}\label{fig:1loop}
\end{figure}

The one-loop amplitude of four Cartan gluons can be written as\footnote{
The perturbative expansion of the amplitude of massless Cartan gluons takes the form
\ie
{\cal A} = g_{YM}^4 {\cal A}^{1-loop} + g_{YM}^6 {\cal A}^{2-loop} + \dotsb + g_{YM}^{2+2L} {\cal A}^{L-loop} + \dotsb.
\fe
}
\ie
\mathcal{A}^{1-loop}(1,2,3,4)= -s_{12}s_{14} \mathcal{A}^{tree}(1,2,3,4)
\left[\mathcal{A}_{1234}^{1-loop} (s_{12},s_{14}) + (2\leftrightarrow 3)+(3\leftrightarrow 4)\right]
\fe
where
\ie
\mathcal{A}_{1234}^{1-loop}(s_{12},s_{14})= \sum_{i,j}\prod_{a=1}^4(v_a^i- v_a^j) \, I^{1-loop}_4(s_{12},s_{14},m_{ij}).
\fe
Here $I^{1-loop}_4(s_{12},s_{14},m_{ij})$ is the scalar box integral (Figure \ref{fig:1loop})\footnote{
In contrast to the more common convention in the scattering amplitude literature (for example (\cite{Bern:2012uf}) where the mostly minus  signature is used and  $s = (p_1+p_2)^2$,  here we work in the mostly plus  signature and define $s = -(p_1+p_2)^2$.  Hence when comparing the two, the Mandelstam variables are the same, but we differ in the definition of the scalar box integrals by factors of $i$ from Wick rotating $d\ell_0$ and minus signs from the propagator $1/p^2$.
}\label{footnote1}
\ie
&I^{1-loop}_4(s_{12},s_{14},m_{ij})\\
&~~~~~~~~
= \int {d^6\ell \over (2\pi)^6} {1\over
(\ell^2+ m_{ij}^2 ) (( \ell+p_1)^2 +m_{ij}^2) (( \ell+p_1+p_2)^2+m_{ij}^2) ((\ell-p_4)^2+m_{ij}^2)
}.
\fe
$m_{ij}$ is the mass of the $W$-boson with gauge indices $(ij)$, and $v_a^j$ is the polarization vector for the external Cartan gluons.

We hope to expand $I^{1-loop}_4$ to $s^2/r^6$ order. It is straightforward to show that
\ie
&I_4^{1-loop}(s_{12},s_{14},m_{ij})\Big|_{s^2\over r^6}
&={1\over 161280 \pi^3 m_{ij}^6}(3 s_{12}^2+3s_{12}s_{13} +2s_{13}^2),
\fe
where we have made the following replacements in the integrand:
\ie
&\ell \cdot p_i \, \ell\cdot p_j \rightarrow {1\over 6}\ell^2\, p_i \cdot p_j = -{1\over 12} \ell^2 s_{ij},\\
&( \ell\cdot p_i)( \ell\cdot p_j)( \ell\cdot p_k)( \ell\cdot p_m ) \rightarrow  {1\over 192}(\ell^2)^2
( s_{ij}s_{km} + s_{ik} s_{jm} + s_{im} s_{jk}).
\fe

Summing up with $\mathcal{A}^{1-loop}_{1324}$ and $\mathcal{A}^{1-loop}_{1243}$, we obtain the order $s^2/r^6$ term  for the full one-loop amplitude
\ie
&\mathcal{A}^{1-loop}(1,2,3,4)\Big|_{s^2\over r^6}
= - 
s_{12}s_{14}\mathcal{A}^{tree}(1,2,3,4)\sum_{i\neq j}\prod_{a=1}^4 (v_a^i -v_a^j) \times {s_{12}^2+s_{13}^2+s_{14}^2\over 46080\pi^3m_{ij}^6}.
\fe

\subsection{Two-loop}

\begin{figure}[tb]
\begin{center}
\includegraphics[scale=0.9]{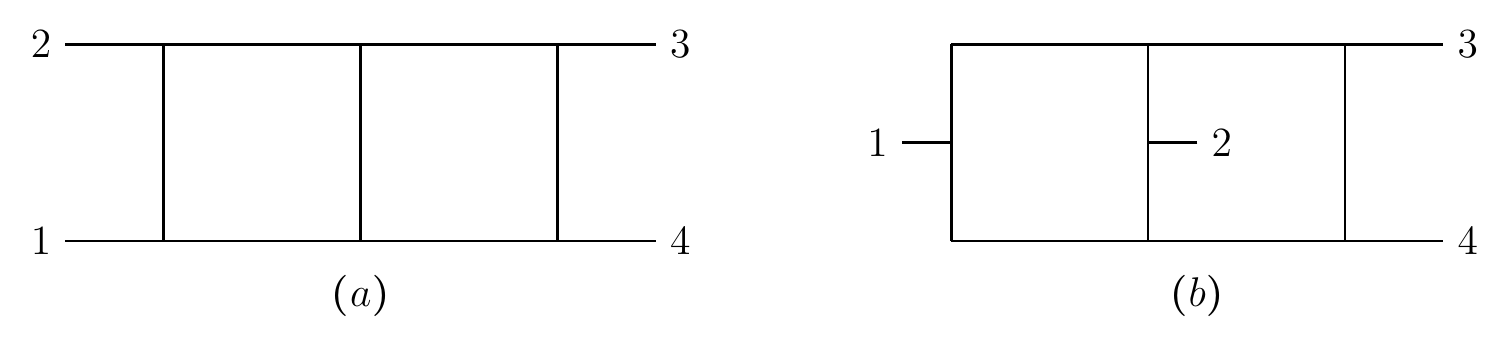}
\end{center}
\caption{In (a), the planar 2-loop scalar integral. In (b), the non-planar 2-loop scalar integral.}\label{fig:2loop}
\end{figure}

The full two-loop amplitude is given by
\ie
&\mathcal{A}^{2-loop}(1,2,3,4)=- s_{12}s_{14} \mathcal{A}^{tree}(1,2,3,4)\\
&\times \left[
s_{12} (\mathcal{A}_{1234}^{2-loop,P} +\mathcal{A}^{2-loop,P}_{3421} +\mathcal{A}^{2-loop,NP}_{1234} 
+\mathcal{A}^{2-loop,NP}_{3421})
+(\text{cyclic in }2,3,4)
\right].
\fe
Let us start with the planar contribution,
\ie
&\mathcal{A}^{2-loop,P}_{1234}\\
&~~~~~= \sum_{i,j,\ell,m,n,r}I_4^{2-loop,P}(m_{ij},m_{\ell m},m_{nr}) 
(\delta_{jn}\delta_{r\ell}\delta_{mi} - 
\delta_{j\ell}\delta_{mn}\delta_{ri})^2
\prod_{a=1,2} (v_a^i -v_a^j)
\prod_{a=3,4} (v_a^m -v_a^\ell),
\fe
where $I_4^{2-loop,P}$ is the planar scalar two-loop integral (Figure \ref{fig:2loop}(a)),
\ie
&I_4^{2-loop,P}(m_{ij}, m_{\ell i}, m_{j\ell}) = \int {d^6\ell_1 \over(2\pi)^6} {d^6\ell_2 \over(2\pi)^6}
{1\over 
(\ell_1^2 +m_{ij}^2 ) ((\ell_1+p_2)^2 +m_{ij}^2)( (\ell_1+p_1+p_2)^2+m_{ij}^2)
}\\
&~~~\times{1\over
(\ell_2^2 +m_{\ell i}^2)((\ell_2+p_3)^2+m_{\ell i }^2)( (\ell_2-p_1-p_2)^2+m_{\ell i}^2) ((\ell_1+\ell_2)^2+m_{j\ell}^2)
}.
\fe
The order $s^2/r^4$ terms in $\mathcal{A}^{2-loop}(1,2,3,4)$ correspond to the $s/r^4$ terms in $I_4^{2-loop,P}$, which can be  computed straightforwardly,
\ie
&I_4^{2-loop,P}\Big|_{s\over r^4}(m_{ij}, m_{\ell i}, m_{j\ell}) 
= \int {d^6\ell_1 \over(2\pi)^6} {d^6\ell_2 \over(2\pi)^6} {1\over (\ell_1^2 +m_{ij}^2)^3 (\ell_2+m_{\ell i}^2)^3( (\ell_1+\ell_2)^2+m_{j\ell}^2)}\\
&~~~~~~~~\times \left[s_{12}\left(
{1\over \ell_1^2+m_{ij}^2}
+{1\over \ell_2^2 +m_{\ell i}^2}
- {\ell_1^2 \over (\ell_1^2+m_{ij}^2)^2}
-{\ell_2^2\over (\ell_2^2+m_{\ell i}^2)^2}
+{4\ell_1\cdot \ell_2 \over 3 (\ell_1^2 +m_{ij}^2)(\ell_2^2+m_{\ell i}^2)}
\right)
\right.\\
&~~~~~~~~\left.
- s_{14}\,{\ell_1\cdot \ell_2\over 3 (\ell_1^2+m_{ij}^2)(\ell_2^2+m_{\ell i}^2)}
\right].
\fe

Moving on to the non-planar diagram,
\ie
&\mathcal{A}^{2-loop,NP}_{1234} 
= \sum_{i,j,\ell,m,n,r}I_4^{2-loop,NP}(m_{ij},m_{\ell m},m_{nr}) 
(\delta_{jn}\delta_{r\ell}\delta_{mi} - 
\delta_{j\ell}\delta_{mn}\delta_{ri})^2
\\
&\hspace{3in} \times
 (v_1^i -v_1^j)(v_2^r-v_2^n)
\prod_{a=3,4} (v_a^m -v_a^\ell),
\fe
where $I^{2-loop,NP}_4$ is the non-planar scalar two-loop integral (Figure \ref{fig:2loop}(b)),
\ie
&I^{2-loop,NP}_4 (m_{ij},m_{\ell i} ,m_{j\ell})
=\int {d^6\ell_1\over (2\pi)^6} {d^6\ell_2\over (2\pi)^6}
{1\over
(\ell_1^2+m_{ij}^2) ((\ell_1+p_1)^2 +m_{ij}^2)
(\ell_2^2+m_{\ell i}^2) ((\ell_2+p_4)^2 + m_{\ell i}^2)}\\
&\times
{1\over
((\ell_2-p_1-p_2)^2+m_{\ell i }^2)
((\ell_1+\ell_2-p_2)^2+m_{j\ell}^2)
((\ell_1+\ell_2)^2+m_{j\ell}^2)
}.
\fe
As in the planar case, we are interested in the $s/r^4$ term in $I_4^{2-loop,NP}$. This can be computed straightforwardly,
\ie
&I^{2-loop,NP}_4 (m_{ij},m_{\ell i} ,m_{j\ell})\Big|_{s\over r^4}
=\int {d^6\ell_1\over (2\pi)^6} {d^6\ell_2\over (2\pi)^6}
{1\over
(\ell_1^2+m_{ij}^2)^2
(\ell_2^2+m_{\ell i}^2)^3 ((\ell_1+\ell_2)^2 + m_{j\ell }^2)^2}\\
&\times
\left[
s_{12}\left(
-{\ell_2^2 \over (\ell_2^2+m_{\ell i}^2)^2}
+{1\over \ell_2^2+m_{\ell i}^2}
+{\ell_1\cdot \ell_2\over 3(\ell_1^2+m_{ ij}^2)(\ell_2^2+m_{\ell i}^2)}
+{\ell_1^2+\ell_1\cdot \ell_2\over 3(\ell_1^2+m_{ij}^2)((\ell_1+\ell_2)^2+m_{j \ell}^2)}
\right.\right.\\
&\left.\left.
-{2\ell_1\cdot \ell_2 +2\ell_2^2 \over 3(\ell_2^2+m_{\ell i}^2)((\ell_1+\ell_2)^2+m_{j\ell}^2)}
\right)
-s_{14}\left(
{\ell_1\cdot\ell_2\over 3(\ell_1^2+m_{ij}^2)(\ell_2^2+m_{\ell i}^2)}
+{\ell_1\cdot \ell_2 +\ell_2^2\over 3(\ell_2^2+m_{\ell i}^2)((\ell_1+\ell_2)^2+m_{j\ell}^2)}
\right)
\right].
\fe

\subsection{Three-loop}
\begin{figure}[htb]
\begin{center}
\includegraphics[scale=1.1]{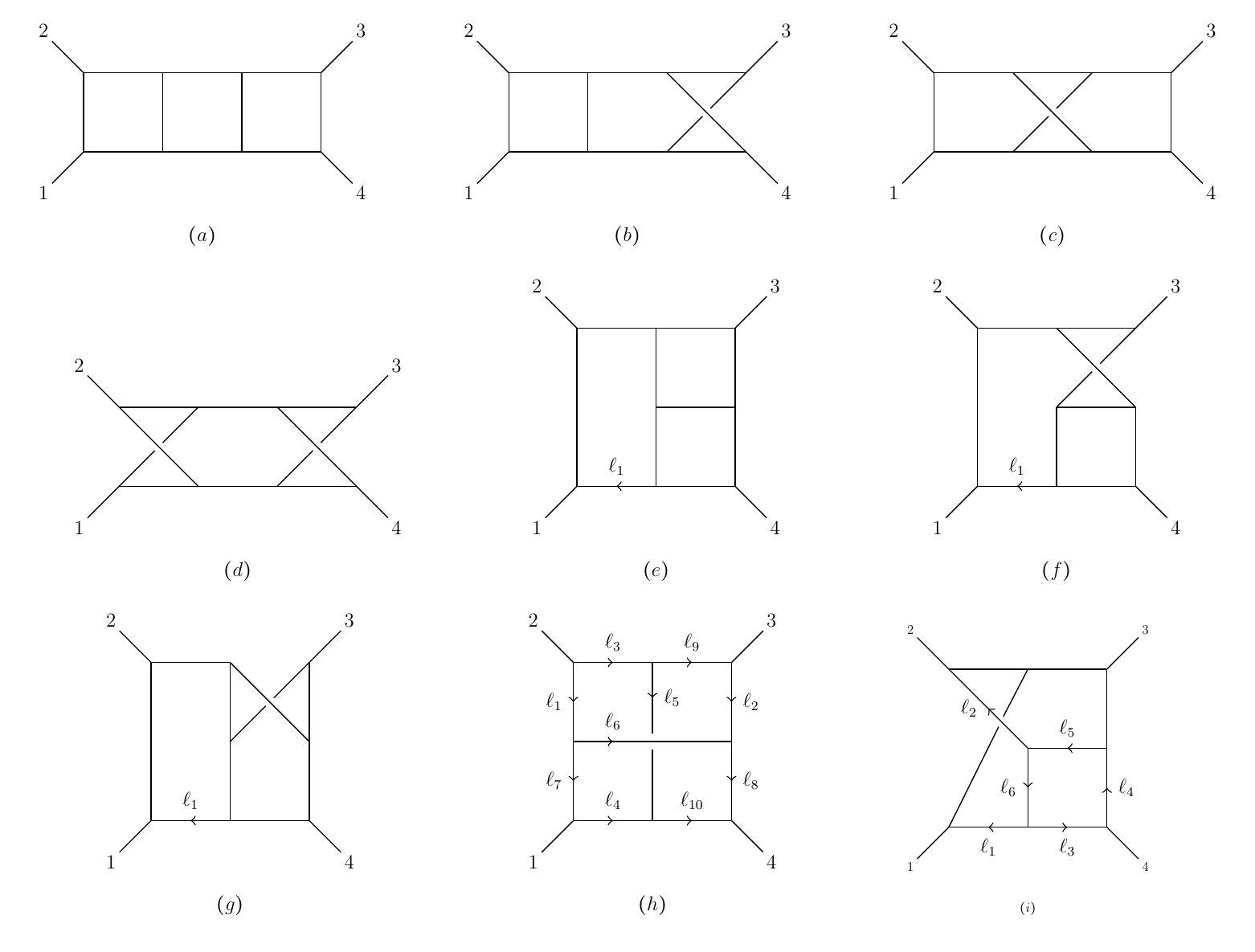}
\end{center}
\caption{The nine 3-loop scalar integrals $I^{(x)}(s_{12},s_{14})$.}\label{fig:3loop}
\label{fig:nine}
\end{figure}

The full three-loop amplitude is given by
\ie
& A^{3-loop}(1,2,3,4)=- s_{12}s_{14}\cA^{tree}(1,2,3,4)
\\
&\times {1\over 4} \sum_{S_4}\left[
\cA^{(a)}_{1234}+\cA^{(b)}_{1234}+{1\over 2}\cA^{(c)}_{1234}+{1\over 4}\cA^{(d)}_{1234}+2\cA^{(e)}_{1234}+2\cA^{(f)}_{1234}+4\cA^{(g)}_{1234}+{1\over 2}\cA^{(h)}_{1234}+2\cA^{(i)}_{1234}
\right]
\fe
where we have summed over contributions from individual diagrams in Figure \ref{fig:3loop}  and permutations of external legs. The coefficients in front of $\mathcal{A}^{(x)}_{1234}$ combined with the overall $1 / 4$ are the symmetry factors. The numerators for the scalar integrals in Figure \ref{fig:3loop} are given in Table \ref{table:numerator}.\footnote{In contrast to the convention in \cite{Bern:2007hh} where the external momenta are all outgoing, our external momenta are all ingoing. Furthermore, as mentioned before, the momentum square $p^2$ differs by a sign due to different conventions on the signature, while the Mandelstam variables are the same. 

Moreover since we consider $W$-bosons propagating through the loops, the loop momenta $\ell_i$ (not all independent) in the expressions of Table \ref{table:numerator}  are taken to be higher dimensional with their extra components constrained by the mass of the propagating particle. These will be made explicit in the expressions for the full scalar integrals below.
  }

\hspace{1in}

\begin{table}[h!]
\begin{center}
\begin{tabular}{|c|c|}
\hline
\text{Integral }$I^{(x)}$ &\text{Numerator Factor}\\
\hline
(a)-(d)&$s_{12}^2$\\
\hline
(e)-(g)& $-s_{12}(\ell_1-p_4)^2$\\
\hline
(h) & $ -s_{12}(\ell_1+\ell_2)^2 - s_{14} (\ell_3+\ell_4)^2 +s_{12} \ell_5^2+s_{14}\ell_6^2 -s_{12}s_{14}$\\
\hline
(i) & $ -s_{12} (\ell_4-\ell_6)^2 +s_{14} (\ell_3-\ell_5)^2+{1\over 3}(s_{12}-s_{14} )\ell_2^2$\\
\hline
\end{tabular}
\end{center}
\caption{The numerator factors in the scalar box integrals in Figure \ref{fig:3loop}. In this table, we omit the $W$-boson mass square $m^2$ term associated to each $(\ell+p)^2$ factor in the numerator. We later restore these factors in the explicit expressions for $\mathcal{A}_{1234}$ below.}\label{table:numerator}
\end{table}

In below we will listed the contribution from each of the nine graphs, with external lines restricted to Cartan gluons, and with the appropriate $W$-boson mass assignments in the internal propagators. The scalar loop integral will then be expanded in powers of external momenta, or in terms of the Mandelstam variables $s,t,u$. At order $s$, while some of the loop integrals are subject to UV divergence, these divergences cancel in the full 3-loop amplitude of Cartan gluons. For the purpose of extracting $D^4 F^4$ effective coupling in the Coulomb effective action, we will expand the scalar integrals to $s^2$ order. Below we will also list these expanded expressions, which can then be evaluated numerically using FIESTA program.

\bigskip

\noindent {\bf Diagram (a)} gives, including color factors,
\ie
{\cal A}^{(a)}_{1234} &= 2 \sum_{i,j,\ell,m} I_a(m_{ij}, m_{i\ell}, m_{im}, m_{j\ell}, m_{\ell m}) \prod_{a=1,2} (v^i_a-v^j_a) \prod_{a=3,4} (v^i_a-v^m_a)
\\
&~~~ + 2  \sum_{i,j,\ell,m}  I_a(m_{ij}, m_{i\ell}, m_{\ell m}, m_{j\ell}, m_{im})\prod_{a=1,2} (v^i_a-v^j_a) \prod_{a=3,4} (v^m_a-v^\ell_a)
\\
& ~~~+ 4 \sum_{i,j} I_a(m_{ij}, 0,m_{ij},m_{ij},m_{ij}) \prod_{a=1}^4 (v^i_a-v^j_a)
\fe
where the scalar integral is 
\ie
&I_a(m_{ij}, m_{i\ell}, m_{im}, m_{j\ell}, m_{\ell m})
\\
&=  s_{12}^2 \int {d^6\ell_1\over (2\pi)^6} {d^6\ell_2\over (2\pi)^6} {d^6\ell_3\over (2\pi)^6} {1\over (\ell_1^2+m_{ij}^2) ((\ell_1+p_1)^2+m_{ij}^2)((\ell_1+p_1+p_2)^2+m_{ij}^2)}
\\
&\times {1\over (\ell_2^2+m_{im}^2)((\ell_2+p_4)^2+m_{im}^2)((\ell_2+p_3+p_4)^2+m_{im}^2) (\ell_3^2+m_{i\ell}^2)((\ell_3+p_1+p_2)^2+m_{i\ell}^2)}
\\
&\times {1\over ((\ell_1-\ell_3)^2+m_{j\ell}^2) ((\ell_2+\ell_3)^2+m_{\ell m}^2)}.
\fe
Before proceeding, let's introduce some shorthand notation,
\ie
  d{\bf L}\equiv {d^6\ell_1\over (2\pi)^6} {d^6\ell_2\over (2\pi)^6} {d^6\ell_3\over (2\pi)^6}
\fe
and
\ie
&\Delta_{i_1 i_2\dots\overline i_a \dots  i_n |p}\equiv (   \ell_{i_1}+  \ell_{i_2}\dots -  \ell_{i_a}\dots+\ell_{i_n })^2+m_p^2
\\
&\tau_{i_1 i_2\dots\overline i_a \dots i_n  ,j_1 j_2\dots\overline j_b \dots j_p }\equiv
  (   \ell_{i_1}+  \ell_{i_2}\dots -  \ell_{i_a}\dots+\ell_{i_n })
\cdot (   \ell_{j_1}+  \ell_{j_2}\dots -  \ell_{j_b}\dots+\ell_{i_n }).
\fe
Expanding in external momenta and extracting the order $s^2$ terms, we have
\ie
\left.I_a(m_1,m_2,m_3,m_4,m_5) \right|_{s^2\over r^2}=s_{12} ^2\int {d{\bf L}\over \Delta_{1|1}^3\Delta_{2|3}^3\Delta_{3|2}^2\Delta_{1\overline 3|4}\Delta_{23|5}}.
\fe
Note that by power counting the loop integral scales like $m_W^{-2} \sim r^{-2}$.


\noindent {\bf Diagram (b)} gives
\ie
{\cal A}^{(b)}_{1234} &= -2\sum_{i,j,\ell,m} I_b(m_{ij}, m_{i\ell}, m_{j\ell},m_{\ell m},m_{im}) \prod_{a=1,2} (v^i_a-v^j_a) (v_3^m-v_3^\ell) (v_4^i-v_4^m)
\\
&~~~-2\sum_{i,j,\ell,m} I_b(m_{ij}, m_{i\ell}, m_{j\ell}, m_{im},m_{\ell m}) \prod_{a=1,2} (v^i_a-v^j_a) (v_4^m-v_4^\ell) (v_3^i-v_3^m)
\\
&~~~ + 4 \sum_{i,j} I_b(m_{ij}, 0, m_{ij}, m_{ij}, m_{ij}) \prod_{a=1}^4 (v^i_a-v^j_a)
,
\fe
where
\ie
& I_b(m_{ij}, m_{i\ell}, m_{j\ell},m_{\ell m},m_{im}) 
\\
&= s_{12}^2 \int {d^6\ell_1\over (2\pi)^6} {d^6\ell_2\over (2\pi)^6} {d^6\ell_3\over (2\pi)^6} {1\over (\ell_1^2+m_{ij}^2) ((\ell_1+p_1)^2+m_{ij}^2)((\ell_1+p_1+p_2)^2+m_{ij}^2)((\ell_1-\ell_3)^2+m_{j\ell}^2)}
\\
&~~~\times {1\over (\ell_2^2+m_{im}^2)((\ell_2+p_4)^2+m_{im}^2)((\ell_2+\ell_3-p_3)^2+m_{\ell m}^2) ((\ell_2+\ell_3)^2+m_{\ell m}^2) }
\\
&~~~\times{1\over(\ell_3^2+m_{i\ell}^2) ((\ell_3+p_1+p_2)^2+m_{i\ell}^2)}.
\fe
%
%
Expanding in external momenta, we have
\ie
\left.I_b(m_1,m_2,m_3,m_4,m_5) \right|_{s^2\over r^2}=s_{12}^2\int {d{\bf L}\over \Delta_{1|1}^3\Delta_{1\overline 3| 3}\Delta_{2|5}^2\Delta_{23|4}^2\Delta_{3|2}^2}.
\fe

\noindent {\bf Diagram (c)} gives
\ie
{\cal A}^{(c)}_{1234} &= 2 \sum_{i,j,\ell,m} I_c(m_{ij}, m_{im}, m_{\ell m}, m_{j\ell}, m_{jm}, m_{i\ell}) \prod_{a=1,2} (v_a^i-v_a^j) \prod_{a=3,4} (v_a^\ell - v_a^m)
\\
&~~~ + 2 \sum_{i,j}  \big[ I_c(m_{ij}, m_{ij}, m_{ij},m_{ij}, 0,0) + I_c(m_{ij},0,m_{ij},0,m_{ij},m_{ij}) \big] \prod_{a=1}^4 (v_a^i-v_a^j),
\fe
where
\ie
&  I_c(m_{ij}, m_{im}, m_{\ell m}, m_{j\ell}, m_{jm}, m_{i\ell})  
\\
&= s_{12}^2 \int {d^6\ell_1\over (2\pi)^6} {d^6\ell_2\over (2\pi)^6} {d^6\ell_3\over (2\pi)^6} {1\over (\ell_1^2+m_{ij}^2) ((\ell_1+p_1)^2+m_{ij}^2)((\ell_1+p_1+p_2)^2+m_{ij}^2)}
\\
&~~~\times {1\over (\ell_2^2+m_{\ell m}^2)((\ell_2+p_4)^2+m_{\ell m}^2)((\ell_2+p_3+p_4)^2+m_{\ell m}^2) (\ell_3^2+m_{j\ell}^2)((\ell_1+p_1+p_2-\ell_2-\ell_3)^2+m_{im}^2)}
\\
&~~~\times {1\over ((\ell_1-\ell_3)^2+m_{i\ell}^2) ((\ell_2+\ell_3)^2+m_{jm}^2)}.
\fe
Expanding in external momenta,
\ie
\left.I_c(m_1,m_2,m_3,m_4,m_5,m_6) \right|_{s^2\over r^2}=s_{12} ^2\int {d{\bf L}\over \Delta_{1|1}^3\Delta_{2|3}^3\Delta_{3|4} \Delta_{1\overline 2\overline3|2}\Delta_{1\overline 3|6}\Delta_{23|5}}.
\fe

\noindent {\bf Diagram (d)} gives 
\ie
{\cal A}^{(d)}_{1234} &= 2\sum_{i,j,\ell,m} I_d(m_{ij},m_{jm}, m_{\ell m}, m_{i\ell}, m_{im}) (v_1^i-v_1^j) (v_2^j-v_2^m) (v_3^\ell-v_3^m) (v_4^i-v_4^\ell)
\\
&~~~ + 2\sum_{i,j,\ell,m} I_d(m_{im}, m_{ij}, m_{\ell m}, m_{j\ell}, m_{jm}) (v_1^m-v_1^i) (v_2^i-v_2^j) (v_3^m-v_3^\ell) (v_4^\ell-v_4^j)
\\
&~~~ + 4\sum_{i,j} I_d(m_{ij},m_{ij},m_{ij},m_{ij},0) \prod_{a=1}^4 (v_a^i-v_a^j)
\\
\fe
where
\ie
& I_d(m_{ij},m_{jm}, m_{\ell m}, m_{i\ell}, m_{im}) 
\\
&= s_{12}^2\int {d^6\ell_1\over (2\pi)^6} {d^6\ell_2\over (2\pi)^6} {d^6\ell_3\over (2\pi)^6} {1\over (\ell_1^2+m_{ij}^2) ((\ell_1+p_1)^2+m_{ij}^2)((\ell_3-\ell_1)^2+m_{jm}^2)((\ell_3-\ell_1+p_2)^2+m_{jm}^2)}
\\
&~~~\times {1\over (\ell_2^2+m_{i\ell}^2)((\ell_2+p_4)^2+m_{i\ell}^2)((\ell_2+\ell_3-p_3)^2+m_{\ell m}^2) ((\ell_2+\ell_3)^2+m_{\ell m}^2)}
\\
&~~~\times{1\over (\ell_3^2+m_{im}^2) ((\ell_3+p_1+p_2)^2+m_{im}^2)}.
\fe
Expanding in external momenta
\ie
\left.I_d(m_1,m_2,m_3,m_4,m_5) \right|_{s^2\over r^2}=s_{12}^2\int {d{\bf L}\over \Delta_{1|1}^2\Delta_{1\overline 3| 2}^2\Delta_{2|4}^2\Delta_{23|3}^2\Delta_{3|5}^2}.
\fe

\noindent {\bf Diagram (e)} gives
\ie
{\cal A}^{(e)}_{1234} &= 2\sum_{i,j,\ell, m} I_e(m_{ij}, m_{i\ell}, m_{im}, m_{jm}, m_{j\ell}, m_{\ell m}) \prod_{a=1,2}(v_a^i-v_a^j) (v_3^i-v_3^\ell) (v_4^i-v_4^m)
\\
&~~~ - 2 \sum_{i,j} I_e(m_{ij},m_{ij},m_{ij},0,0,0) \prod_{a=1}^4(v_a^i-v_a^j) ,
\fe
where
\ie
& I_e(m_{ij}, m_{i\ell}, m_{im}, m_{jm}, m_{j\ell}, m_{\ell m})
\\
& =  -s_{12}\int {d^6\ell_1\over (2\pi)^6} {d^6\ell_2\over (2\pi)^6} {d^6\ell_3\over (2\pi)^6} {(\ell_1-p_4)^2 +m_{ij}^2\over (\ell_1^2+m_{ij}^2)((\ell_1+p_1)^2+m_{ij}^2)((\ell_1+p_1+p_2)^2+m_{ij}^2)}
\\
&~~~ \times {1\over (\ell_2^2+m_{i\ell}^2)((\ell_2+p_3)^2+m_{i\ell}^2) (\ell_3^2+m_{im}^2)((\ell_3+p_4)^2+m_{im}^2)}
\\
&~~~ \times {1\over ((\ell_3-\ell_1+p_4)^2+m_{jm}^2)((\ell_1-\ell_2+p_1+p_2)^2+m_{j\ell}^2) ((\ell_2-\ell_3+p_3)^2+m_{\ell m}^2)}.
\fe

Expanding in external momenta, and after some simplification of the loop integrals, we have
\ie
&\left.I_e(m_1,m_2,m_3,m_4,m_5,m_6) \right|_{s^2\over r^2}
= {s_{12} \over 3}\int {d{\bf L} \over \Delta_{1|1}^2\Delta_{2|2}^2\Delta_{3|3}^2\Delta_{1\overline 3|4}\Delta_{1\overline 2|5}
\Delta_{2\overline 3|6}}  \Bigg[-{3s_{12}\over \Delta_{1|1}}-{3s_{12}\over \Delta_{1\overline 2|5} }
\\
&\quad
+
{(s_{14}+2s_{12})\tau_{1 ,1 }\over \Delta^2_{1 |1}}
+{2s_{12} \tau_{1,1\overline 2}\over\Delta_{1|1} \Delta_{1\overline 2 |5}}
+{\tau_{1,2} (- s_{12}-s_{14})\over \Delta_{1|1}\Delta_{2|2}}
+{\tau_{1,3}(s_{14}-s_{12})\over  \Delta_{1|1}\Delta_{3|3}}
+{\tau_{1,3\overline 1}(s_{14}-s_{12})\over  \Delta_{1|1}\Delta_{1\overline 3|4}}
\\
&\quad
+{\tau_{1,2\overline 3}(- s_{12}-s_{14})\over  \Delta_{1|1}\Delta_{2\overline 3|6}}
+{\tau_{2,3} s_{12}\over  \Delta_{3|3}\Delta_{2|2}}
+{\tau_{2,3\overline 1} s_{12} \over \Delta_{2|2}\Delta_{1\overline 3|4}}
-{\tau_{2,1\overline 2} s_{12}\over  \Delta_{2|2}\Delta_{1\overline 2|5}}
-{\tau_{3,1\overline 2} s_{12}\over  \Delta_{3|3}\Delta_{1\overline 2|5}}
+{\tau_{3,2\overline 3}s_{12} \over  \Delta_{3|3}\Delta_{2\overline 3|6}}
\\
&\quad
+{\tau_{1\overline 3,1\overline 2} s_{12}\over  \Delta_{1\overline 3|4}\Delta_{1\overline 2|5}}
+{\tau_{3\overline 1,2\overline 3} s_{12} \over \Delta_{3\overline 1|4}\Delta_{2\overline 3|6}}
-{\tau_{1\overline 2,2\overline 3}  s_{12} \over  \Delta_{1\overline 2|5}\Delta_{2\overline 3|6}}
+{2 \tau_{1\overline 2,1\overline 2}s_{12} \over \Delta_{1\overline 2|5}^2}
\Bigg].
\fe

\noindent {\bf Diagram (f)} gives
\ie
{\cal A}^{(f)}_{1234} &= -2 \sum_{i,j,\ell,m} I_f(m_{ij}, m_{j\ell}, m_{im}, m_{jm}, m_{\ell m}, m_{i\ell}) \prod_{a=1,2}(v_a^i-v_a^j) (v_3^\ell-v_3^j) (v_4^i-v_4^m)
\\
& ~~~ -2 \sum_{i,j} I_f(m_{ij},m_{ij}, m_{ij}, 0,m_{ij}, 0)\prod_{a=1}^4 (v_a^i-v_a^j),
\fe
where
\ie
&I_f(m_{ij}, m_{j\ell}, m_{im}, m_{jm}, m_{\ell m}, m_{i\ell})
\\
&= - s_{12} \int {d^6\ell_1\over (2\pi)^6} {d^6\ell_2\over (2\pi)^6} {d^6\ell_3\over (2\pi)^6} {(\ell_1-p_4)^2+m_{ij}^2\over (\ell_1^2+m_{ij}^2) ((\ell_1+p_1)^2+m_{ij}^2)((\ell_1+p_1+p_2)^2+m_{ij}^2) (\ell_2^2+m_{im}^2)}
\\
&~~~ \times {1\over((\ell_2+p_4)^2+m_{im}^2)((\ell_1-\ell_3+p_1+p_2)^2+m_{j\ell}^2) ((\ell_1-\ell_3-p_4)^2+m_{j\ell}^2) ((\ell_2+\ell_3+p_4)^2+m_{\ell m}^2)  }
\\
&~~~ \times {1\over ((\ell_1+\ell_2)^2+m_{jm}^2) (\ell_3^2+m_{i\ell}^2)}.
\fe
Expanding in external momenta, we have
\ie
&\left.I_f(m_1,m_2,m_3,m_4,m_5,m_6) \right|_{s^2\over r^2}=
 {s_{12}\over 3}\int {d{\bf L}\over \Delta_{1|1}^2\Delta_{2|3}^2\Delta_{1\overline 3|2}^2\Delta_{23|5}\Delta_{3|6}\Delta_{12|4}} 
 \\
 &\quad\times \Bigg[
-{3s_{12}\over \Delta_{1|1}}-{3s_{12}\over\Delta_{1\overline 3|2}}
+{\tau_{1,1}(s_{14}+2s_{12})\over \Delta_{1|1}^2}
 + {\tau_{1,2}(s_{14}-s_{12})\over  \Delta_{1|1} \Delta_{2|3}} 
 - {\tau_{2,1\overline 3} s_{12} \over  \Delta_{2|3} \Delta_{1\overline 3|2}}
 + {\tau_{1,1\overline 3} (3s_{12}-s_{14})\over  \Delta_{1|1} \Delta_{1\overline 3|2}}
 \\
 &\quad
 + {\tau_{1,23} (s_{14}-s_{12}) \over  \Delta_{1|1} \Delta_{23|5}} 
 - {\tau_{1\overline 3,23}s_{12} \over  \Delta_{1\overline 3|2} \Delta_{23|5}}
  +{3 \tau_{1\overline 3,1\overline 3} s_{12} \over  \Delta_{1\overline 3|2}^2}
 \Bigg] .
\fe

\noindent {\bf Diagram (g)} gives 
\ie
{\cal A}^{(g)}_{1234}&= - 2\sum_{i,j} I_g(m_{ij}, m_{ij}, m_{ij}, 0, m_{ij},0) \prod_{a=1}^4 (v_a^i-v_a^j)
\\
&~~~ - 2 \sum_{i,j,\ell,m} I_g(m_{ij}, m_{\ell m}, m_{im}, m_{jm}, m_{j\ell}, m_{i\ell}) \prod_{a=1,2} (v_a^i-v_a^j) (v_3^\ell-v_3^m) (v_4^i-v_4^m),
\fe
where
\ie
& I_g(m_1, m_2, m_3, m_4, m_5, m_6)
\\
&=- s_{12} \int {d^6\ell_1\over (2\pi)^6} {d^6\ell_2\over (2\pi)^6} {d^6\ell_3\over (2\pi)^6} {(\ell_1-p_4)^2+m_1^2\over (\ell_1^2+m_1^2) ((\ell_1+p_1)^2+m_1^2)((\ell_1+p_1+p_2)^2+m_1^2) (\ell_2^2+m_3^2)}
\\
&~~~ \times {1\over((\ell_2+p_4)^2+m_3^2)((\ell_1-\ell_3+p_1+p_2)^2+m_5^2) ((\ell_2+\ell_3+p_3+p_4)^2+m_2^2) ((\ell_2+\ell_3+p_4)^2+m_2^2)  }
\\
&~~~\times {1\over ((\ell_1+\ell_2)^2+m_4^2)( \ell_3^2+m_6^2)}.
\fe
Expanding in external momenta, we have
\ie
&\left.I_g(m_1,m_2,m_3,m_4,m_5,m_6) \right|_{s^2\over r^2}= {s_{12}\over 3}\int {d{\bf L}\over \Delta_{1|1}^2\Delta_{2|3}^2\Delta_{1\overline 3|5}\Delta_{23|2}^2\Delta_{3|6}\Delta_{12|4}}
\\&\quad\times
\Bigg[
-{3s_{12}\over \Delta_{1|1}}-{3s_{12}\over \Delta_{1\overline 3|5}}-{3s_{12}\over \Delta_{23|2}}+
{\tau_{1,1}(s_{14}+2s_{12})\over \Delta_{1|1}^2}+ {\tau_{1,2}(s_{14}-s_{12})\over \Delta_{1|1}\Delta_{2|3}} 
+ {2\tau_{1,1\overline 3}s_{12} \over\Delta_{1|1}\Delta_{1\overline 3|5}} 
- {\tau_{2,1\overline 3} s_{12}\over \Delta_{2|3}\Delta_{1\overline 3|5}}
\\
&\quad
+ {\tau_{1,23}(-3s_{12}+s_{14})\over \Delta_{1|1}\Delta_{23|2}}
+ {\tau_{2,23} s_{12}\over \Delta_{2|3}\Delta_{23|2}}
- {3\tau_{1\overline 3,23} s_{12} \over \Delta_{1\overline 3|5}\Delta_{23|2}}
+{ 3  \tau_{23,23} s_{12} \over   \Delta_{23|2}^2}
+{ 2  \tau_{1\overline 3,1\overline 3}s_{12} \over   \Delta_{1\overline 3|5}^2}
\Bigg].
\fe

\noindent {\bf Diagram (h)} gives
\ie
{\cal A}^{(h)}_{1234} &= 2 \sum_{i,j} I_h(m_{ij}, m_{ij}, m_{ij}, m_{ij}, 0,0) \prod_{a=1}^4(v_a^i-v_a^j)
\\
&~~~ + 2 \sum_{i,j,\ell,m} I_h(m_{ij}, m_{i\ell}, m_{\ell m}, m_{jm}, m_{j\ell}, m_{im}) (v_1^i-v_1^j) (v_2^i-v_2^\ell) (v_3^m-v_3^\ell) (v_4^m-v_4^j),
\fe
where
\ie
& I_h(m_{ij}, m_{i\ell}, m_{\ell m}, m_{jm}, m_{j\ell}, m_{im})
\\
&= \int {d^6\ell_1\over (2\pi)^6} {d^6\ell_2\over (2\pi)^6} {d^6\ell_3\over (2\pi)^6} {- s_{12}((\ell_1+\ell_2)^2-(\ell_1+\ell_2-p_2-p_3)^2) - s_{14} ((\ell_3-p_1-p_2)^2-\ell_3^2) - s_{12} s_{14} \over (\ell_1^2+m_{i\ell}^2) ((\ell_1-p_2)^2+m_{i\ell}^2)(\ell_2^2+m_{\ell m}^2) ((\ell_2-p_3)^2+m_{\ell m}^2)}
\\
&~~~\times {1\over((\ell_1-\ell_3)^2+m_{ij}^2)((\ell_1-\ell_3+p_1)^2+m_{ij}^2) ((\ell_2+\ell_3)^2+m_{jm}^2) ((\ell_2+\ell_3+p_4)^2+m_{jm}^2)  }
\\
&~~~\times {1\over ((\ell_1+\ell_2-p_2-p_3)^2+m_{im}^2) (\ell_3^2+m_{j\ell}^2)}.
\fe
Expanding in external momenta, we have
\ie
&\left.I_h(m_1,m_2,m_3,m_4,m_5,m_6) \right|_{s^2\over r^2}=
-{s_{12}s_{23}\over 3}\int {d{\bf L}\over \Delta_{1|2}^2\Delta_{2|3}^2\Delta_{1\overline 3|1}^2\Delta_{23|4}^2\Delta_{12|6}\Delta_{3|5}} 
\\
&\quad\times
\Bigg[
3
+{\tau_{3\overline 1\overline 2,1}\over \Delta_{1|2} }
- {\tau_{312,2}     \over \Delta_{2| 3}}
- {\tau_{312,1\overline 3}  \over \Delta_{1\overline 3| 1}} 
+ {\tau_{3\overline 1\overline 2,23}  \over\Delta_{23| 4}} 
- {2\tau_{12,12} \over\Delta_{12| 6}}
\Bigg].
\fe

\noindent {\bf Diagram (i)} gives
\ie
{\cal A}^{(i)}_{1234} &= -2 \sum_{i,j,\ell,m} I_i(m_{ij}, m_{j\ell}, m_{i\ell}, m_{im}, m_{jm}, m_{\ell m}) (v_1^i-v_1^j)(v_2^j-v_2^\ell) (v_3^i-v_3^\ell) (v_4^i-v_4^m),
\fe
where
\ie
& I_i(m_{ij}, m_{j\ell}, m_{i\ell}, m_{im}, m_{jm}, m_{\ell m})
\\
&= \int {d^6\ell_1\over (2\pi)^6} {d^6\ell_2\over (2\pi)^6} {d^6\ell_3\over (2\pi)^6} 
{-s_{12}((\ell_1-p_4)^2+m_{ij}^2) + s_{14} ((\ell_1+\ell_2)^2+m_{i\ell}^2) + {1\over 3} (s_{12}-s_{14})(\ell_2^2+m_{j\ell}^2)\over (\ell_1^2+m_{ij}^2) ((\ell_1+p_1)^2+m_{ij}^2) (\ell_2^2+m_{j\ell}^2) ((\ell_2+p_2)^2+m_{j\ell}^2)}
\\
&~~~ \times {1\over ((\ell_1+\ell_2+p_1+p_2)^2+m_{i\ell}^2) ((\ell_1+\ell_2-p_4)^2+m_{i\ell}^2)
(\ell_3^2+m_{im}^2)((\ell_3+p_4)^2+m_{im}^2)}
\\
&~~~\times {1\over ((\ell_1+\ell_3)^2+m_{jm}^2) ((\ell_1+\ell_2+\ell_3)^2+m_{\ell m}^2)}.
\fe
Expanding in external momenta, we have
\ie
&\left.I_i(m_1,m_2,m_3,m_4,m_5,m_6) \right|_{s^2\over r^2}=
{1\over 3 } \int {d{\bf L}
\over 
\Delta_{1|1}^2\Delta_{2|2}^2\Delta_{12|3}^2\Delta_{3|4}^2\Delta_{13|5}\Delta_{123|6} }
\\
&\quad\times\Bigg[s_{12}\Bigg(
{\tau_{1,1} s_{14} \over\Delta_{1|1}}-{\tau_{1,2} (s_{12}+s_{14})\over \Delta_{2|2}}
-{\tau_{12,1}  s_{12}\over \Delta_{12|3}}\Bigg)
\\
&\quad
-\Bigg(-s_{12}\Delta_{1|1} + s_{14} \Delta_{12|3} + {1\over 3} (s_{12}-s_{14})\Delta_{2|2}\Bigg)
\Bigg(-{3\over\Delta_{12|3} }
+
{\tau_{1,2} s_{12} \over \Delta_{1|1}\Delta_{2|2}}
+{\tau_{1,12}  (s_{12}-s_{14}) \over \Delta_{1|1}\Delta_{12|3}}
\\&\quad
+{\tau_{1,3} s_{14} \over \Delta_{1|1}\Delta_{3|4}}
+{\tau_{2,12}(2s_{12}+s_{14}) \over  \Delta_{2|2}\Delta_{12|3}}
-{\tau_{2,3}(s_{12}+s_{14})\over \Delta_{2|2}\Delta_{3|4}}
+{3\tau_{12,12} s_{12} \over \Delta_{12|3}^2}
-{\tau_{12,3} s_{12}\over \Delta_{12|3}\Delta_{3|4}}
\Bigg)
\Bigg].
\fe

Note that the above expressions for the scalar loop integrals expanded in external momenta to order $s^2$ do not always exhibit symmetries of the graphs in a manifest way. In the numerical evaluation of the loop integrals, verification of these symmetries is a basic and useful consistency check.

\paragraph{Results for 6D SYM in the Coulomb Phase}
To make contact with the consideration of 6D SYM in Section~\ref{sec:sym}, we set the mass of the $W$-boson with gauge indices $(ij)$ to be
\ie
m_{ij} = 2r \Big| \sin {\pi (i-j)\over k}\Big|,
\fe
and the polarization vector for the external Cartan gluons to be
\ie
v^j_a =\omega^{(j-1)n_a},~~~j=1,\cdots,k,
\fe
where $\omega= e^{2\pi i/k}$.  For the four Cartan gluon scattering of interest, 
\ie
n_1=n_2=\ell+1,~~~n_3=n_4=k-(\ell+1)
\fe
with values $\ell=0,1,\cdots, k-2$.  


The partial amplitudes and full amplitudes for each case are listed in the tables below.  The quantity listed is the three-loop contribution to $D^4 F^4$ normalized by the one-loop $F^4$ amplitude
\ie
&\mathcal{A}^{1-loop}(1,2,3,4)\Big|_{s^2\over r^6}
\\
&=- s_{12}s_{14}\mathcal{A}^{tree}(1,2,3,4)\, 
{s_{12}^2+s_{13}^2+s_{14}^2\over r^6}
{k\over184320} \sum_{L=1}^{k-1}
{\sin^2 {\pi L (\ell+1)\over k}\sin^2 {\pi L (k-\ell-1)\over k}\over \sin^6 {\pi L\over k}}.
\fe
In the notation of Section~\ref{sec:sym}, this quantity is $C^3_S (s^2 + t^2 + u^2) + C^3_A s^2$.

\begin{itemize}

\item  $\mathbf{k=2, ~\ell = 0}$

\begin{center}
\begin{tabular}{|c|c|c|}
\hline
diagram & $g_{YM}^4 \mathcal{A}^{3-loop} / \mathcal{A}^{1-loop}$ & symmetry factor
\\\hline\hline
(a) & $6.603600 (s^2+t^2+u^2)$ & 4
\\
(b) & $3.2071994 (s^2+t^2+u^2)$ & 4
\\
(c) & $2.6718092 (s^2+t^2+u^2)$ & 8
\\
(d) & $2.4143983 (s^2+t^2+u^2)$ & 16
\\
(e) & $0$ & 2
\\
(f) & $0.55684116 (s^2+t^2+u^2)$ & 2
\\
(g) & $0.54568714 (s^2+t^2+u^2)$ & 1
\\
(h) & $0.089231678 (s^2+t^2+u^2)$ & 8
\\
(i) & $0$ & 2
\\\hline\hline
total & $3.772838 (s^2+t^2+u^2)$ &
\\\hline
\end{tabular}
\end{center}

\item  $\mathbf{k = 3, ~\ell = 0}:$

\begin{center}
\begin{tabular}{|c|c|c|}
\hline
diagram & $g_{YM}^4 \mathcal{A}^{3-loop} / \mathcal{A}^{1-loop}$ & symmetry factor
\\\hline\hline
(a) & $14.39876 (s^2+t^2+u^2) - 10.376120 s^2$ & 4
\\
(b) & $5.976425 (s^2+t^2+u^2) - 4.223506 s^2$ & 4
\\
(c) & $5.1697610 (s^2+t^2+u^2) - 3.7469277 s^2$ & 8
\\
(d) & $3.8144749 (s^2+t^2+u^2) - 1.2321663 s^2$ & 16
\\
(e) & $-0.56439858 (s^2+t^2+u^2) + 0.42112441 s^2$ & 2
\\
(f) & $0.68831287 (s^2+t^2+u^2) + 0.37394094 s^2$ & 2
\\
(g) & $1.0393916 (s^2+t^2+u^2) - 0.73705051 s^2$ & 1
\\
(h) & $0.17584295 (s^2+t^2+u^2) - 0.12991690 s^2$ & 8
\\
(i) & $-0.030527986 (s^2+t^2+u^2) + 0.091583958 s^2$ & 2
\\\hline\hline
total & $7.086485 (s^2+t^2+u^2) - 4.505248 s^2$ &
\\\hline
\end{tabular}
\end{center}

\item  $\mathbf{k = 4, ~\ell = 0}:$

\begin{center}
\begin{tabular}{|c|c|c|}
\hline
diagram & $g_{YM}^4 \mathcal{A}^{3-loop} / \mathcal{A}^{1-loop}$ & symmetry factor
\\\hline\hline
(a) & $25.02079 (s^2+t^2+u^2) - 20.545614 s^2$ & 4
\\
(b) & $9.696932 (s^2+t^2+u^2) - 8.164688 s^2$ & 4
\\
(c) & $8.584892 (s^2+t^2+u^2) - 7.224038 s^2$ & 8
\\
(d) & $5.8421205 (s^2+t^2+u^2) - 2.2545368 s^2$ & 16
\\
(e) & $-1.336350 (s^2+t^2+u^2) + 0.918643 s^2$ & 2
\\
(f) & $0.9180370 (s^2+t^2+u^2) + 0.824604 s^2$ & 2
\\
(g) & $1.7063675 (s^2+t^2+u^2) - 1.4403125 s^2$ & 1
\\
(h) & $0.28983076 (s^2+t^2+u^2) -0.24273063 s^2$ & 8
\\
(i) & $-0.06256573 (s^2+t^2+u^2) + 0.1816815 s^2$ & 2
\\\hline\hline
total & $11.619831 (s^2+t^2+u^2) - 8.729678 s^2$ &
\\\hline
\end{tabular}
\end{center}

\item  $\mathbf{k = 4, ~\ell = 1}:$

\begin{center}
\begin{tabular}{|c|c|c|}
\hline
diagram & $g_{YM}^4 \mathcal{A}^{3-loop} / \mathcal{A}^{1-loop}$ & symmetry factor
\\\hline\hline
(a) & $17.16058 (s^2+t^2+u^2)$ & 4
\\
(b) & $6.703913 (s^2+t^2+u^2)$ & 4
\\
(c) & $6.131683 (s^2+t^2+u^2)$ & 8
\\
(d) & $4.8779369 (s^2+t^2+u^2)$ & 16
\\
(e) & $-0.762594 (s^2+t^2+u^2)$ & 2
\\
(f) & $1.252848 (s^2+t^2+u^2)$ & 2
\\
(g) & $1.2121707 (s^2+t^2+u^2)$ & 1
\\
(h) & $0.21142504 (s^2+t^2+u^2)$ & 8
\\
(i) & $0$ & 2
\\\hline\hline
total & $8.521180 (s^2+t^2+u^2)$ &
\\\hline
\end{tabular}
\end{center}

\item  $\mathbf{k = 5, ~\ell = 0}:$

\begin{center}
\begin{tabular}{|c|c|c|}
\hline
diagram & $g_{YM}^4 \mathcal{A}^{3-loop} / \mathcal{A}^{1-loop}$ & symmetry factor
\\\hline\hline
(a) & $38.51941 (s^2+t^2+u^2) - 33.12847 s^2$ & 4
\\
(b) & $14.416872 (s^2+t^2+u^2) - 13.026020 s^2$ & 4
\\
(c) & $12.923744 (s^2+t^2+u^2) - 11.564527 s^2$ & 8
\\
(d) & $8.4375334 (s^2+t^2+u^2) - 3.5261340 s^2$ & 16
\\
(e) & $-2.318423 (s^2+t^2+u^2) + 1.526873 s^2$ & 2
\\
(f) & $1.2298478 (s^2+t^2+u^2) + 1.375730 s^2$ & 2
\\
(g) & $2.552601 (s^2+t^2+u^2) - 2.301272 s^2$ & 1
\\
(h) & $0.43320460 (s^2+t^2+u^2) - 0.38238309 s^2$ & 8
\\
(i) & $-0.1008116 (s^2+t^2+u^2) + 0.292874 s^2$ & 2
\\\hline\hline
total & $ 17.38894 (s^2+t^2+u^2) - 13.955903 s^2$ &
\\\hline
\end{tabular}
\end{center}

\item  $\mathbf{k = 5, ~\ell = 1}:$

\begin{center}
\begin{tabular}{|c|c|c|}
\hline
diagram & $g_{YM}^4 \mathcal{A}^{3-loop} / \mathcal{A}^{1-loop}$ & symmetry factor
\\\hline\hline
(a) & $27.87823 (s^2+t^2+u^2) - 20.43163 s^2$ & 4
\\
(b) & $10.203492 (s^2+t^2+u^2) - 7.860821 s^2$ & 4
\\
(c) & $9.645323 (s^2+t^2+u^2) - 7.922393 s^2$ & 8
\\
(d) & $6.151710 (s^2+t^2+u^2) - 1.9295194 s^2$ & 16
\\
(e) & $-1.538466 (s^2+t^2+u^2) + 0.7303561 s^2$ & 2
\\
(f) & $1.308312 (s^2+t^2+u^2) + 0.645134 s^2$ & 2
\\
(g) & $1.879007 (s^2+t^2+u^2) - 1.451108 s^2$ & 1
\\
(h) & $0.32813257 (s^2+t^2+u^2) - 0.26805768 s^2$ & 8
\\
(i) & $-0.0512934 (s^2+t^2+u^2) + 0.1579127 s^2$ & 2
\\\hline\hline
total & $ 12.88988 (s^2+t^2+u^2) - 8.901921 s^2$ &
\\\hline
\end{tabular}
\end{center}

\end{itemize}

\subsection{Four-loop}

The result of \cite{Bern:2012uf} for the 4-loop 4-point amplitude of maximal $SU(k)$ SYM in $D=6-2\epsilon$ dimensions is 
\ie
\label{A4L}
& {\cal A}^{4-loop}(1,2,3,4) = (st {\cal A}_{tree}(1,2,3,4)) {e^{-4\C\epsilon}\over (4\pi)^{12-4\epsilon}} k \Bigg\{ \left({\rm Tr}_{12} {\rm Tr}_{34} + {\rm Tr}_{14} {\rm Tr}_{23} + {\rm Tr}_{13} {\rm Tr}_{24}\right)
\\
&~~~~~~~~~\times (s^2+t^2+u^2) \left[ - {k^2+36\zeta_3\over 2\epsilon^2} + {1\over \epsilon} \left( k^2 \left({35\over 18} - {\zeta_3\over 3}\right) + 4\zeta_3 + 9 \zeta_4 + 20 \zeta_5 \right) \right] 
\\
&~~~~~~~~~~~~ - {3\over \epsilon} (k^2 \zeta_3+25\zeta_5) \left( {\rm Tr}_{12} {\rm Tr}_{34} s^2+ {\rm Tr}_{14} {\rm Tr}_{23} t^2 + {\rm Tr}_{13} {\rm Tr}_{24} u^2 \right) \Bigg\}
+ ({\rm single~trace}).
\fe
When restricted to the Cartan gluons, of charge $n_a\in\mathbb{Z}_k$ ($a=1,2,3,4$) with respect to the $\mathbb{Z}_k$ action, the single trace term is always proportional to $(s^2+t^2+u^2) \delta_{\sum n_a}$ ($\delta$ here stands for Kronecker delta modulo $k$). The coefficient will involve $1/\epsilon^2$ and $1/\epsilon$ divergences. These have not been computed explicitly. 

On the other hand, for the double trace terms, we have 
\ie
{\rm Tr}_{ab} =
\begin{cases}
k, & n_a+n_b \equiv 0 \mod k,
\\
0, & \rm{otherwise}.
\end{cases}
\fe
For the amplitude of gluons with $\mathbb{Z}_k$ charge $(n,n,-n,-n)$ ($n=\ell+1$ in our notation), we always have ${\rm Tr}_{13}={\rm Tr}_{14}={\rm Tr}_{23}={\rm Tr}_{24}=k$. ${\rm Tr}_{12}={\rm Tr}_{34}=0$ for $n\not=k/2$, and ${\rm Tr}_{12}={\rm Tr}_{34}=k$ for $n=k/2$. In the case $k=4$, by comparing $\ell=0$ with $\ell=1$, we can separate a contribution from double trace terms only,
\ie
& {\cal A}^{4-loop}_{k=4,\ell=1} - {\cal A}^{4-loop}_{k=4,\ell=0} = (st {\cal A}_{tree}) {e^{-4\C\epsilon}\over (4\pi)^{12-4\epsilon}} 64 \Bigg\{ (s^2+t^2+u^2) 
\\
&~~~~~~\times  \left[ - {16+36\zeta_3\over 2\epsilon^2} + {1\over \epsilon} \left( 16 \left({35\over 18} - {\zeta_3\over 3}\right) + 4\zeta_3 + 9 \zeta_4 + 20 \zeta_5 \right) \right] 
- {3\over \epsilon} (16 \zeta_3+25\zeta_5) s^2 \Bigg\}
\fe
After subtracting off the 4-loop counter-terms, we expect
\ie
& {\cal A}^{4-loop}_{k=4,\ell=1} - {\cal A}^{4-loop}_{k=4,\ell=0} ={ (st {\cal A}_{tree}) \over (4\pi)^{12}} 64 \Bigg\{  (s^2+t^2+u^2) \left[-(8+18 \zeta_3) (8 \ln r)^2 + A\ln r + B\right]
\\
&~~~~~~
+ s^2\cdot 3 (16 \zeta_3+25\zeta_5) (8 \ln r + C) \Bigg\}.
\fe
Here $A$ is a constant that depends on finite shifts of the 3-loop $D^2 {\rm tr}F^4$ counter-term, and $B,C$ are constants that depend on finite shifts of the 4-loop $D^4 {\rm tr} F^4$ and $D^4 {\rm tr}^2 F^4$ counter-terms. They cannot be determined from SYM perturbation theory alone.

In the $n = {k / 2}$ cases, all terms are proportional to $s^2+t^2+u^2$, and we cannot separate the double trace terms from the single trace terms at all. In the $k=3$ and $k=5$ cases, as well as the $k=4,\ell=0$ case, since ${\rm Tr}_{12}={\rm Tr}_{34}=0$, we can determine
\ie
& {\cal A}^{4-loop}_{k,\ell} = {(st {\cal A}_{tree}) \over (4\pi)^{12}} k^3 \Bigg\{  (s^2+t^2+u^2) ({\rm unknown})
- s^2\cdot 3 (k^2 \zeta_3+25\zeta_5) (8 \ln r + C) \Bigg\}.
\fe

\section{Evaluation of the Little String Amplitudes}

In this appendix, we discuss some machinery that went into the numerical evaluation of the double scaled little string theory amplitude (\ref{ALST}).  The conformal block can be written in the form~\cite{Zamolodchikov:1985ie,Zamolodchikov:1987}
\ie
\label{Zamo}
F(\Delta_{i}; \Delta_P | z) &= (16 q)^{P^2} z^{{Q^2\over 4} - \Delta_{1} - \Delta_{2}} (1-z)^{{Q^2\over 4} - \Delta_{1} - \Delta_{3}} \\
& \hspace{.5in} \times \theta_3(q)^{3Q^2 - 4 (\Delta_{1}+\Delta_{2}+\Delta_{3}+\Delta_{4}) } H(\Delta_i; \Delta_P | q),
\fe
where $\Delta_P = {Q^2 \over 4} + P^2$, $z$ is the cross ratio
\ie
z = {z_{12} z_{34}\over z_{14} z_{32}},
\fe
$q$ is the nome of $z$, defined by
\ie
q(z)=e^{\pi i\tau(z)} ,\quad \tau(z) = i {K(1-z)\over K(z)},~~~~ K(z) = {1\over 2} \int_0^1 {dt\over \sqrt{t(1-t) (1-z t) }},
\fe
and $\theta_3$ is the Jacobi theta function defined by
\ie
\theta_3(p) = \sum_{n=-\infty}^\infty p^{n^2}.
\fe
$H$ satisfies Zamolodchikov's recurrence formula~\cite{Zamolodchikov:1985ie,Zamolodchikov:1987},
which allows one to obtain $H$ as a series expansion in $q$. Alternatively, we can compute $F$ as a series expansion in $z$ by computing inner products between Virasoro descendants of the external primary states.  The resulting expression is manifestly a rational function in $c$, $\Delta_i$, and $\Delta_P$.  For this reason the latter brute-force method is more advantageous for obtaining simple analytic expressions, although its computational complexity (with respect to the order of the series in $q$) is much higher than the complexity of the recurrence method.

The conformal block written in the form (\ref{Zamo}) converges much faster than a naive series expansion in $z$, due to the fact that $|q(z)|$ is much smaller than $z$ (note for example that $16 |q(z)| \leq |z|$ and $|q(z)| < 1$ for all $z \in \mathbb C$).  Given an order-$N$ series in $z$, we can rewrite it in the form of (\ref{Zamo}) by performing a variable transformation and then truncate $H$ to order $q^N$.  If we want to integrate $z$ over regions far from the origin, it is crucial that we approximate the conformal block by a truncation of (\ref{Zamo}) instead of a series in $z$.

The Liouville structure constant $C(\A_1, \A_2, \A_3)$ is expressed as ratios of the special function $\Upsilon$, which has an integral representation~\cite{Zamolodchikov:1995aa,Teschner:2001gi}
\ie
\log \Upsilon(x) = \int_0^\infty {dt \over t} \left[ \left( {Q \over 2} - x \right)^2 e^{-t} - { \sinh^2({Q \over 2} - x) {t \over 2} \over \sinh {bt \over 2} \sinh {t \over 2b} } \right]
\fe
that is is convergent for $0 < {\rm Re}\,x < Q$.  For $x$ lying outside this region, $\Upsilon$ can be analytically continued via the shift formulae
\ie
\Upsilon_b(x+b) = \gamma(bx) b^{1-2bx} \Upsilon_b(x), \quad \Upsilon_b(x+{1 / b}) = \gamma({x / b}) b^{{2x \over b} - 1} \Upsilon_b(x),
\fe
where
\ie
\gamma(x) \equiv {\Gamma(x) \over \Gamma(1-x)}.
\fe
When evaluating $\Upsilon$ numerically, the oscillatory behavior of the second term at large $t$ must be taken care of by stripping out an exponential integral function
\ie
\int_{t_0}^\infty {dt \over 4t} { e^{({Q \over 2}-x) t} \over \sinh {bt \over 2} \sinh {t \over 2b} }
&= {\rm E}_1(x t_0) + \int_{t_0}^\infty {dt \over 4t} { ( e^{-bt} + e^{-t \over b} - e^{-Q t} ) e^{({Q\over 2}-x) t} \over \sinh {bt \over 2} \sinh {t \over 2b} }.
\fe
To obtain the Liouville four-point function, we then integrate over the Liouville momentum $P$ of the intermediate state.  This integral is performed by a simple Riemann sum.

Finally we are in place to evaluate the integral with respect to the cross ratio $z$.  We break the integral over the complex plane into six regions.  These regions are mapped to each other under the $S^3$ action generated by $z \to 1-z$ and $z \to {1 / z}$.  A fundamental region near the origin
\ie
I: \quad |z| \leq 1, \quad {\rm Re}\,z < {1 \over 2}
\fe
is chosen and the integrals over the other regions are mapped to Region $I$ using crossing symmetry of the four-point functions.  In Region $I$, $|z|$ is bounded by 1, and $|q|$ by $0.066$, thus with the conformal block expressed in the form of (\ref{Zamo}), even if $H$ is truncated to $q^6$ order, we still have at least $10^{-7}$ precision for $F$!

%

\bibliography{lstrefs,6dsymrefs} 
\bibliographystyle{JHEP}

\end{document}